\documentclass[11pt,a4paper,oneside]{article}
\usepackage[top=3cm, bottom=3cm, left=2cm, right=2cm]{geometry}
\usepackage[english]{babel}
\usepackage[utf8x]{inputenc}
\usepackage[T1]{fontenc}

\usepackage{amsthm,amsmath,amssymb,mathrsfs,dsfont,mathtools}

\usepackage{bm}
\usepackage{graphicx}
\usepackage[colorlinks=true, allcolors=blue]{hyperref}
\usepackage{booktabs,subcaption}
\usepackage{hyperref}
\usepackage[all]{nowidow}

\usepackage{setspace}
\usepackage[ruled,norelsize,vlined]{algorithm2e}
\usepackage{tikz} 
\usetikzlibrary{shapes.geometric, arrows}
\tikzstyle{process} = [rectangle, rounded corners, minimum width=4cm, minimum height=1.2cm, text centered, draw=black]
\tikzstyle{arrow} = [thick,->,>=stealth]

\usepackage{sectsty}
\allsectionsfont{\centering \normalfont\scshape} 
\usepackage[round]{natbib}

\usepackage{authblk}
\title{A generalized Bayes framework for probabilistic clustering}
\author[1]{Tommaso Rigon}
\author[1]{Amy H. Herring}
\author[1]{David B. Dunson}
\affil[1]{Department of Statistical Science, Duke University, Durham, U.S.A.}

\date{}
\newtheorem{theorem}{Theorem}

\newtheorem{lemma}{Lemma}
\newtheorem{proposition}{Proposition}

\theoremstyle{definition}
\newtheorem{definition}{Definition}
\newtheorem{remark}{Remark}
\newtheorem{example}{Example}

\makeatletter
\def\env@cases{%
  \let\@ifnextchar\new@ifnextchar
  \left\lbrace
  \def\arraystretch{0.8}%
  \array{@{}l@{\quad}l@{}}}
\makeatother

\newcommand \dd  { \,\textup d}   

\begin{document}
\maketitle

\begin{abstract}
Loss-based clustering methods, such as k-means and its variants, are standard tools for finding groups in data. However, the lack of quantification of uncertainty in the estimated clusters is a disadvantage.  Model-based clustering based on mixture models provides an alternative, but such methods face computational problems and large sensitivity to the choice of kernel.  This article proposes a generalized Bayes framework that bridges between these two paradigms through the use of Gibbs posteriors.  In conducting Bayesian updating, the log likelihood is replaced by a loss function for clustering, leading to a rich family of clustering methods.  The Gibbs posterior represents a coherent updating of Bayesian beliefs without needing to specify a likelihood for the data, and can be used for characterizing uncertainty in clustering.  We consider losses based on Bregman divergence and pairwise similarities, and develop efficient deterministic algorithms for point estimation along with sampling algorithms for uncertainty quantification.  Several 
existing clustering algorithms, including k-means, can be interpreted as generalized Bayes estimators under our framework, and hence we provide a method of uncertainty quantification for these approaches.
\end{abstract}

\smallskip
{\small \noindent \textbf{Keywords}: Bayesian; Bregman divergence; Gibbs posterior; Loss function; Product partition model; Uncertainty quantification}

\section{Introduction}\label{sec:intro}

\begin{figure}[t]
\includegraphics[width=\textwidth]{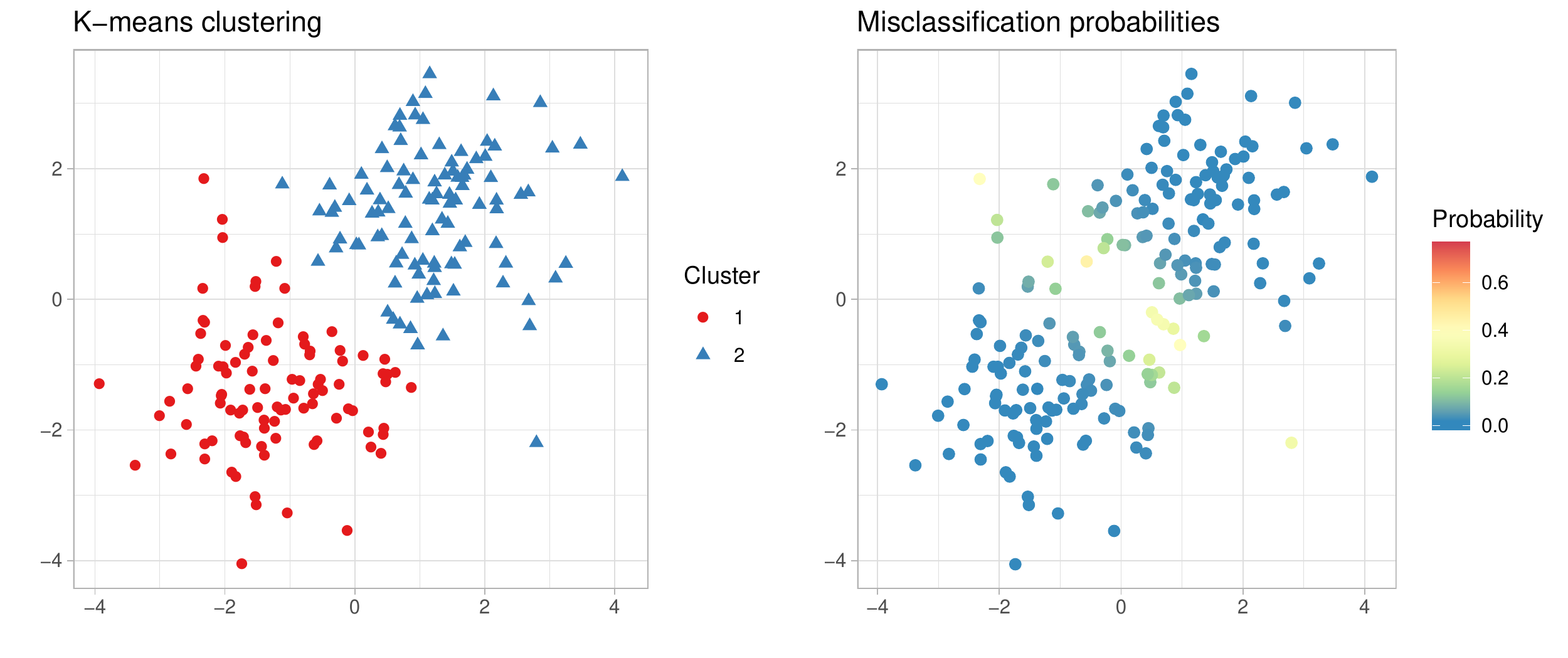}
\caption{Clustering and uncertainty quantification on a toy example with $n = 50$ observations. Left plot: data points, with colors representing the k-means solution. Right plot: generalized Bayes misclassification probabilities associated with k-means. \label{fig:intro}}
\end{figure}

Cluster analysis is a canonical topic in statistical learning, which continues to draw substantial attention.  Particularly popular are ``algorithmic'' approaches such as k-means, its generalizations based on Bregman divergences \citep{Banerjee2005}, and the k-medoids  (\textsc{pam}) algorithm; see e.g. \citet{Kaufman1990}, \cite{Hastie2008} and \citet{Jain2010} for an overview. These methods rely on reasonable heuristics typically aimed at minimizing a specific loss function, and they are widely employed  because they are conceptually simple, computationally efficient (k-means), and/or robust to  moderate perturbations of the data (\textsc{pam}). However, despite their popularity, these approaches lack  uncertainty quantification. As an example, consider the dataset on the left side of Figure~\ref{fig:intro}: although k-means provides a sensible solution, some observations lie at the cluster boundaries. In order to quantify this phenomenon, one could consider the misclassification probability, namely the chance that an observation belongs to one of the other groups. Unfortunately, it is currently unclear how to compute such probabilities for k-means. 

Model-based clustering based on mixtures provides a rigorous inferential foundation that naturally leads to  probabilistic assessments \citep{Fraley2002,Fruhwirth2019}. While conceptually appealing, mixture models face substantial practical difficulties, including computational bottlenecks and sensitivity to misspecification. Posterior inference for Bayesian mixture models is routinely conducted via Markov chain Monte Carlo (\textsc{mcmc}) which, despite remarkable advances \citep[e.g.][]{Stephens2000,Jain2004,Wang2011,Zuanetti2019}, remains computationally expensive and affected by the label-switching issue. In addition, misspecified mixtures may have unreliable and overly complex clustering solutions, thus requiring robust procedures which may increase the computational burden; see e.g. \citet{Rodriguez2014,Miller2019,Lijoi2020} for some recent advances in the Bayesian setting. 

In this paper we aim at bridging the loss- and model-based frameworks through a generalized Bayes approach that provides a probability distribution characterizing uncertainty in clustering without requiring one to specify a likelihood for the data. We do this within a coherent Bayesian framework for updating prior beliefs using information in the data.  Let $\bm{X}$ be the data, $\bm{\theta}$ a parameter of interest and $\pi(\bm{\theta})$ the corresponding prior. Then, one can consider the generalized posterior
\begin{equation}\label{GibbsPosterior}
\pi(\bm{\theta} \mid \lambda, \bm{X}) \propto \pi(\bm{\theta})\exp\{ - \lambda \ell(\bm{\theta}; \bm{X}) \},
\end{equation}
with $\lambda > 0$ and $\ell(\bm{\theta}; \bm{X}) > 0$ an arbitrary loss function. Expression (\ref{GibbsPosterior}) is known as the \emph{Gibbs posterior} \citep{Shawe1997,McAllester1998,Jiang2008}. Standard Bayesian inference is a special case of~\eqref{GibbsPosterior}, occurring when the loss function $\lambda \ell(\bm{\theta}; \bm{X})$ is a negative log-likelihood. Although Gibbs posteriors have been used in a wide variety of settings, it is only recently that they were shown to provide a rational update of beliefs and hence can be regarded as genuine posterior distributions 
\citep{Bissiri2016}.  \citet{Holmes2017} provide an important development in developing methods for choosing $\lambda$.

Despite the increasing theoretical support for equation~\eqref{GibbsPosterior}, to our knowledge the Gibbs posterior framework has not yet been used to define a general methodology for clustering.  There is a small literature on  generalized Bayes methods for clustering, but these approaches tend to lack the link with loss function based methods such as k-means.  Importantly, they cannot be used to address our interest in 
providing uncertainty quantification for k-means and related methods.  A key recent example is the coarsened posterior framework of \citet{Miller2019}, which they use to provide a robust approach for Bayesian model-based clustering.  Refer to \citet{Gorsky2020} for a successful application of this framework. Alternatively, \citet{Duan2019} propose a Bayesian model for a pairwise distance matrix representation of the data.  Early ideas for clustering estimation that combine prior information with general losses appear in \citet{Lau2007}. Also relevant are the algorithms of \citet{Kulis2012, Broderick2013}, which are based on the very different idea of starting with a Bayesian mixture model and taking a small variance limit to define algorithms for clustering. 

In contrast, we propose a very broad Gibbs posterior-based framework for uncertainty quantification in clustering.  A rich class of loss functions for clustering are considered, and 
we extensively study the resulting properties.  We additionally introduce novel algorithms for point estimation and uncertainty quantification, including for k-means and other existing clustering algorithms.  Our framework is also appealing in terms of simplicity and interpretability of tuning parameters.

The paper is organized as follows. In Section~\ref{sec:method} we 
propose a generalized Bayes product partition modeling 
(\textsc{gb-ppm}) framework and discuss properties.  
In Section~\ref{sec:posterior_inference} we propose algorithms for posterior inference.  In Section~\ref{sec:Bregman} we study the use of Bregman divergences within \textsc{gb-ppm}, and develop highly efficient computational algorithms.  In Section~\ref{sec:distance} we consider pairwise dissimilarities, show robustness to contamination, and develop a novel k-dissimilarities algorithm for point estimation.
The above classes are extremely broad, and we discuss important special cases in Section~\ref{sec:models}.  In 
 Section~\ref{sec:illustration} we illustrate the methods through simulation studies and a medical application. Concluding remarks are given in Section~\ref{sec:discussion}.

\section{Generalized Bayes clustering}\label{sec:method}

\subsection{Background and Motivation}\label{sec:background}

Let $\bm{x}_i = (x_{i1},\dots, x_{id})^\intercal$ be a vector of observations on $\mathds{X} \subseteq \mathds{R}^d$, for $i=1,\dots,n$, and let $\bm{X}$ be the collection of all the data points. Moreover, suppose $\bm{C} = \{C_1,\dots,C_K\}$ is a partition of the integers $\{1,\dots,n\}$ into $K = |\bm{C}|$ disjoint sets. Alternatively, the partition can be described through cluster labels $\bm{c} = (c_1,\dots,c_n)$, so that $c_i = k$ for $i=1,\dots,n$ if and only if $i \in C_k$.  In this case, only the induced partition is relevant and the labels on the indicators $\bm{c} = (c_1,\dots,c_n)$ are not of interest. 
Typical Bayesian models for clustering are based on posterior distributions of the form
\begin{equation}\label{mixturemodel}
\pi(\bm{c} \mid \bm{X}) \propto \pi(\bm{c}) \prod_{k=1}^K \left[\int_{\Theta} \prod_{i \in C_k} \pi(\bm{x}_i \mid \bm{\theta}) \pi(\bm{\theta}) \dd \bm{\theta}\right],
\end{equation}
where $\pi(\bm{c})$ is the prior probability of $\bm{c}$, 
$\pi(\bm{x} \mid \bm{\theta})$ is the within-cluster likelihood, 
and $\pi(\bm{\theta})$ is the prior law generating the cluster-specific parameters. 

Although equation~\eqref{mixturemodel} forms the basis for a vast literature on Bayesian clustering, there are key practical problems that arise.  The integral is often not available in closed form, leading to computational complications.  More importantly, the posterior on clustering is very sensitive to the precise specification of 
 $\pi(\bm{x} \mid \bm{\theta})$ and $\pi(\bm{\theta})$. The assumption is that observations are drawn from 
 \begin{equation}\label{latent}
(\bm{x}_i \mid \bm{\theta}_k, c_i = k) \overset{\textup{ind}}{\sim} \pi(\bm{x}_i \mid \bm{\theta}_k),  \qquad i \in C_k, \quad k=1,\dots,K,
\end{equation}
where $\bm{\theta}_k \overset{\textup{iid}}{\sim} \pi(\bm{\theta})$, for $k=1,\dots,K$. There are two critical problems with (\ref{latent}): (i) clustering may be a convenient simplification and there may not actually be distinct groups in the data, and (ii) even if there are, the distribution of the data within each cluster is unlikely to exactly follow the chosen distribution $\pi(\bm{x} \mid \bm{\theta})$. Unfortunately, the posterior on clustering is highly sensitive to both issues, leading to a critical brittleness problem.

\subsection{Generalized Bayes product partition models}

We introduce a class of \emph{generalized Bayes product partition models (\textsc{gb-ppm})} for clustering. This broad family is characterized by a factorized loss $\ell(\bm{c}; \bm{X})$. We consider the case in which the number of clusters $K$ is fixed, e.g. because it is known or has been selected in an exploration phase. We discuss the choice of $K$ in Section~\ref{sec:bissiri}. Let $\bm{X}_k = \{\bm{x}_i : i \in C_k\}$ denote the observations belonging to cluster $C_k$, for $k=1,\dots,K$. Then, \textsc{gb-ppm} models are characterized by losses admitting the factorization
\begin{equation}\label{general_loss}
\ell(\bm{c}; \bm{X}) = \sum_{k=1}^K\sum_{i \in C_k} \mathcal{D}(\bm{x}_i; \bm{X}_k),  \qquad  \bm{c} : |\bm{C}| = K, 
\end{equation}
where $\mathcal{D}(\bm{x}_i; \bm{X}_k) \ge 0$ is a function of $\bm{x}_i$ and $\bm{X}_k$ which quantifies the discrepancy of the $i$th unit from $k$th cluster. As a default, we focus on uniform clustering priors having the form 
\begin{equation}\label{unifprior}
\pi(\bm{c}) = \frac{1}{\mathcal{S}(n,K)}, \qquad \bm{c} : |\bm{C}| = K,
\end{equation}
where $\mathcal{S}(n,K) = 1/K! \sum_{k=0}^K (-1)^{K-k}K!\{(K-k)!k!\}^{-1}k^n$ is the Stirling number of the second kind. Prior~\eqref{unifprior} is uniform over partitions having $K$ components; our framework is easily modified to consider more elaborate clustering priors, but we focus on the uniform case throughout the paper.

\begin{definition}\label{product_partition} Let the loss function $\ell(\bm{c}; \bm{X})$ be as in~\eqref{general_loss} and the prior $\pi(\bm{c})$ be as in~\eqref{unifprior}. The generalized Bayes posterior 
under a \emph{generalized Bayes product partition model (\textsc{gb-ppm})} has the form
\begin{equation*}
\pi(\bm{c} \mid \lambda, \bm{X}) \propto \pi(\bm{c})\prod_{k=1}^K \rho(C_k; \lambda, \bm{X}_k)  \propto \prod_{k=1}^K \exp\left\{ - \lambda \sum_{i \in C_k} \: \mathcal{D}(\bm{x}_i; \bm{X}_k)\right\}, \qquad \bm{c} : |\bm{C}| = K.
\end{equation*}
\end{definition}

The quantity $\pi(\bm{c} \mid \lambda, \bm{X})$ is a well-defined probability mass function, because the normalizing constant is such that $0 < \sum_{\bm{c} : |\bm{C}| = K}\pi(\bm{c}) \prod_{k=1}^K \rho(C_k; \lambda, \bm{X}_k) < \infty$.  Our proposed \textsc{gb-ppm} relates to the existing literature on Bayesian product partition models; refer to
\citet{Quintana2003,Park2010,Muller2011} among others.  Such models use 
Definition~\ref{product_partition} to define a data-based prior for clustering, while inheriting the disadvantages described in Section~\ref{sec:background} in relying on 
(\ref{mixturemodel}) and requiring specification of 
$\pi(\bm{x} \mid \bm{\theta})$ and $\pi(\bm{\theta})$.

A primary contribution of this paper is to use the generalized posterior in Definition~\ref{product_partition} for inference on clustering including uncertainty quantification.  The \textsc{gb-ppm} framework is extremely broad, and we will consider a variety of important subclasses corresponding to different choices of $\mathcal{D}(\bm{x}_i; \bm{X}_k)$.  Perhaps the simplest example is squared error loss as described in Example~\ref{kmeans}.

\begin{example}[k-means loss]\label{kmeans} Let $||\bm{x}||_2^2 = x_1^2 + \cdots + x_d^2$  for any $\bm{x} \in \mathds{X}$. Moreover, let $\bar{\bm{x}}_k$ represent the vector with the arithmetic means of the columns of $\bm{X}_k$, for any $k=1,\dots,K$. Then the loss
\begin{equation}\label{kmeans_loss}
\ell(\bm{c}; \bm{X}) = \sum_{k=1}^K \sum_{i \in C_k} ||\bm{x}_i - \bar{\bm{x}}_k||_2^2,
\end{equation}
defines a \textsc{gb-ppm}. 
\end{example}

\subsection{Decision theoretic justification}\label{sec:bissiri}

We now show that any \textsc{gb-ppm} can be rightfully regarded as a posterior distribution, being a rational update of one's prior beliefs, thus adapting the reasoning of \citet{Bissiri2016} to the clustering setting. In our \textsc{gb-ppm} framework, the target of inference is an optimal and unknown partition $\bm{c}_\textsc{opt}$, defined as the minimizer of an integrated loss function, namely
\begin{equation}\label{copt}
\bm{c}_\textsc{opt} = \arg \min_{\bm{c}} \mathds{E}_{\pi_0}\{ \ell(\bm{c}; \bm{X}) \} = \arg \min_{\bm{c} : |\bm{C}| = K} \sum_{k=1}^K\sum_{i \in C_k} \mathds{E}_{\pi_0}\left\{ \mathcal{D}(\bm{x}_i; \bm{X}_k)\right\},
\end{equation}
where the expectation is taken with respect to the unknown data generating process $\pi_0(\bm{X})$. The term $\mathds{E}_{\pi_0}\{\ell(\bm{c}; \bm{X}) \}$ is sometimes called the \emph{frequentist risk}.  The definition of 
$\bm{c}_\textsc{opt}$ does not require the existence of a true partition of the data.  Even in the absence of truly distinct groups of observations within a dataset, it is often of interest to infer clusters that capture aspects of the data encoded by the loss function.

The target $\bm{c}_\textsc{opt}$ cannot be estimated based on a single dataset without knowledge of $\pi_0(\bm{X})$.  Instead we rely on 
the \textsc{gb-ppm} posterior, which can be viewed as the ``best'' 
conditional distribution for quantifying our subjective beliefs about $\bm{c}_\textsc{opt}$. Let $\mathscr{P}$ be the space of all conditional distributions given the data and let $\mathscr{L}$ be a loss function defined in such a space. For any $\nu_1,\nu_2 \in \mathscr{P}$, the loss $\mathscr{L}$ describes whether $\nu_1$ is a better candidate than $\nu_2$  for representing ones' posterior beliefs about $\bm{c}_\textsc{opt}$. 
Let $\pi(\bm{c} \mid \lambda, \bm{X}) = \arg \min_{\nu \in \mathscr{P}} \mathscr{L}\{\nu(\bm{c})\}$. A reasonable candidate for $\mathscr{L}$ is the loss
\begin{equation}\label{loss_gibbs}
 \mathscr{L}\{\nu(\bm{c})\} =  \lambda \mathds{E}_\nu\left\{\ell(\bm{c}; \bm{X})\right\} + \textsc{kl}\{\nu(\bm{c}) \mid\mid \pi(\bm{c}) \},
\end{equation}
where the expectation is taken with respect to $\nu(\bm{c})$. It is easy to show that the unique minimizer of equation~\eqref{loss_gibbs} is indeed the \textsc{gb-ppm} of Definition~\ref{product_partition}. 

The loss $\mathscr{L}$ in equation~\eqref{loss_gibbs} balances two components: the discrepancy with respect to the observed data $\mathds{E}_\nu\left\{\ell(\bm{c}; \bm{X})\right\}$, and the closeness to the prior $\textsc{kl}\{\nu(\bm{c}) \mid\mid \pi(\bm{c})\}$. The parameter $\lambda > 0$ controls the weight assigned to the former component. To clarify the relationship between the minimization problem in~\eqref{copt} and the loss $\mathscr{L}$ of equation~\eqref{loss_gibbs},  consider the following extreme cases. As $\lambda 
\rightarrow 0$ the closeness to the data is not penalized and therefore one obtains $\pi(\bm{c} \mid 0, \bm{X})  = \pi(\bm{c})$. As $\lambda \rightarrow \infty$ the effect of the prior is negligible and the Gibbs posterior collapses to a point mass~$\delta_{\hat{\bm{c}}_\textsc{opt}}$, where $\hat{\bm{c}}_\textsc{opt} = \arg \min_{\bm{c}} \ell(\bm{c}; \bm{X})$ is the so called \emph{empirical risk minimizer}, i.e. the empirical counterpart of $\bm{c}_\textsc{opt}$. Hence, the \textsc{gb-ppm} posterior combines the empirical version of $\bm{c}_\textsc{opt}$ with one's prior beliefs.

The above discussion suggests that a \textsc{gb-ppm} should be regarded as a standard Bayesian posterior, with a slight but crucial distinction: the target of inference is not anymore some latent partition as in equation~\eqref{mixturemodel}, but the more general object $\bm{c}_\textsc{opt}$. Therefore, within this context,  the number of clusters $K$ should be regarded as the resolution at which we want to partition the observations rather than some attribute of the data that we should estimate.  Hence, we believe that the number of clusters $K$ should be subjectively specified as part of the loss $\ell(\bm{c}; \bm{X})$, rather than inferred from the data. Alternatively, $K$ could be selected in an exploration phase using standard strategies such as the ``elbow'' rule. Refer also to Section~\ref{sec:illustration} for practical examples. 

\section{Posterior inference}\label{sec:posterior_inference}
\subsection{Point estimation and Gibbs sampling}

We develop a set of strategies for posterior inference for \textsc{gb-ppm}s.  We find that implementation is typically easier and much more efficient compared with mixture models.  In particular, very efficient algorithms for point estimation are available.  For uncertainty quantification, we rely on Gibbs sampling strategies that are straightforward and are often characterized by good mixing.   

We first consider point estimation. Although several alternatives exist \citep[e.g.][]{Medvedovic2004, Lau2007,Fritsch2009,Wade2018}, maximum a posteriori (\textsc{map}) estimation provides a particularly natural and simple choice.  In particular, we let 
\begin{equation}
\hat{\bm{c}}_{\textsc{map}} = \arg \max_{\bm{c}} \pi(\bm{c} \mid \lambda, \bm{X}).
\end{equation}
In a \textsc{gb-ppm} the $\hat{\bm{c}}_{\textsc{map}}$ can be equivalently obtained by minimizing the loss function $\ell(\bm{c}; \bm{X})$ over the space of partitions having $K$ components. Hence, $\hat{\bm{c}}_\textsc{map}$ coincides with the empirical risk minimizer $\hat{\bm{c}}_\textsc{opt}$. The proof of the following Proposition is trivial, but the practical implications are important.

\begin{proposition}\label{map} Let $\pi(\bm{c} \mid \lambda, \bm{X})$ be a \textsc{gb-ppm}. Then,
\begin{equation*}
\hat{\bm{c}}_{\textsc{map}} = \arg \min_{\bm{c} \: : \: |\bm{C}| = K} \sum_{k=1}^K\sum_{i \in C_k} \mathcal{D}(\bm{x}_i; \bm{X}_k).
\end{equation*}
\end{proposition}
The estimate $\hat{\bm{c}}_\textsc{map}$ in Proposition~\ref{map} does not depend on $\lambda$.  It turns out that $\hat{\bm{c}}_{\textsc{map}}$ can be efficiently computed in several cases, as it coincides with the solution of well known clustering methods. This links the algorithmic approaches to our framework, as the following example clarifies.  

\begin{example}[k-means loss, cont'd] In a \textsc{gb-ppm} with loss function~\eqref{kmeans_loss} the \textsc{map} estimate is
\begin{equation*}
\hat{\bm{c}}_{\textsc{map}} = \arg \min_{\bm{c} \: : \: |\bm{C}| = K} \sum_{k=1}^K\sum_{i \in C_k}  ||\bm{x}_i - \bar{\bm{x}}_k||_2^2, \qquad \bm{c} : |\bm{C}| = K,
\end{equation*}
which corresponds to the k-means estimator \citep[e.g.][]{Jain2010}. 
\end{example}

For non-\textsc{map} point estimation and uncertainty quantification, we instead rely on Gibbs sampling.  Let $\bm{c}_{-i} = (c_1,\dots,c_{i-1},c_{i+1},\dots,c_n)$ be the collection of cluster indicators without the $i$th unit, and let $\{C_{1,-i},\dots,C_{K,-i}\}$ be the associated partition, data points $\bm{X}_{k,-i} = \{\bm{x}_i : i \in C_{k,-i}\}$, and cohesion functions $\rho(C_{k,-i}; \lambda, \bm{X}_{k,-i})$. In Gibbs sampling we cyclically re-allocate the indicators $c_i$ by sampling from their full conditionals.  In a \textsc{gb-ppm} we focus on partitions $\bm{c}$ such that $|\bm{C}| = K$, implying that if the $i$th unit is the only element in a cluster, it can not be re-allocated. 

\begin{theorem}\label{GibbsSampling} Let $\pi(\bm{c} \mid \lambda, \bm{X})$ be a \textsc{gb-ppm}. Then, the conditional distribution of $c_i$ given $\bm{c}_{-i}$ is
\begin{equation*}
\mathds{P}(c_i =k \mid \bm{c}_{-i}, \lambda, \bm{X}) \propto \frac{\rho(C_k; \lambda, \bm{X}_k)}{\rho(C_{k,-i}; \lambda, \bm{X}_{k,-i})} = \exp\left\{ - \lambda \left[\sum_{i' \in C_k} \mathcal{D}(\bm{x}_{i'}; \bm{X}_k) - \sum_{i' \in C_{k,-i}} \mathcal{D}(\bm{x}_{i'}; \bm{X}_{k,-i}) \right] \right\},
\end{equation*}
for $k = 1,\dots,K$ and for any partition $\bm{c} : |\bm{C}| = K$. \end{theorem}

This stochastic allocation has an intuitive interpretation, since the involved probabilities are ratios of cohesion functions. In other words, the $i$th unit is likely to be allocated in the $k$th cluster if the cohesion of the newly created cluster $\rho(C_k; \lambda, \bm{X}_k)$ is higher than the old cohesion $\rho(C_{k,-i}; \lambda, \bm{X}_{k,-i})$.

\begin{remark} Gibbs sampling strategies based on the re-allocations of the labels are not novel \citep[e.g.][]{Escobar1995}, and they are generally affected by slow mixing e.g. when transitioning from $K$ to $K'$ clusters. However, this issue does not occur in \textsc{gb-ppm}s because the number of groups is fixed.  Indeed, we are considering a simplified sampling problem characterized by fewer local optima. 
\end{remark}

\subsection{Clustering validation and uncertainty quantification}\label{sec:uq}

A cluster solution is especially useful when it can be qualitatively validated and interpreted. To this purpose, we adapt the notion of \emph{centroid} and \emph{medoid} to our context. Secondly, we emphasize that uncertainty quantification can be very helpful in practice, because it provides a more complete description of the clustering problem. To illustrate this crucial aspect, we review several inferential quantities that one may want to consider in a cluster analysis.


Let $\bm{m}_k \in \mathds{X}$ denote the centroid of the $k$th cluster.
Each term $\rho(C_k; \lambda, \bm{X}_k) = \prod_{i \in C_k} \rho(\bm{x}_i; \lambda, \bm{X}_k)$ can be written as the product of unit-specific contributions, each representing the cohesion of the $i$th unit to the $k$th cluster. Hence, given a specific partition $C_1,\dots,C_K$ and recalling that $\bm{X}_k = \{\bm{x}_i : i \in C_k\}$, we define the associated centroids $\bm{m}_1,\dots,\bm{m}_K$ as 
\begin{equation*}
\bm{m}_k = \arg \max_{\bm{x} \in \mathds{X}} \rho(\bm{x}; \lambda, \bm{X}_k) = \arg \min_{\bm{x} \in \mathds{X}}\mathcal{D}(\bm{x}; \bm{X}_k), \qquad k=1,\dots,K,
\end{equation*}
that is, the centroid is defined as the value having the maximal cohesion within the $k$th cluster. 
For instance, in the k-means Example~\ref{kmeans} one has that $\bm{m}_k = \arg \min_{\bm{x} \in \mathds{X}} ||\bm{x} - \bar{\bm{x}}_k||_2^2$, implying that $\bm{m}_k = \bar{\bm{x}}_k$, thus recovering the usual definition of centroid. A closely related alternative is the so called medoid, which is a representative data point $\bm{x}_{i^*_k}$ of the $k$th cluster. The associated indices are obtained as
\begin{equation*}
i^*_k = \arg \max_{i \in C_k} \rho(\bm{x}_i; \lambda, \bm{X}_k) = \arg \min_{i \in C_k}\mathcal{D}(\bm{x}_i; \bm{X}_k), \qquad k=1,\dots,K.
\end{equation*}
The medoids can be always computed because they involve a search over a finite set of points. We will show in Section~\ref{sec:Bregman} that the centroids can be also easily obtained in most cases. 

We now build on existing tools for uncertainty quantification in clustering problems. In principle, it would be interesting to consider the probabilities $\mathds{P}(c_i = k \mid \lambda, \bm{X})$, but these are affected by the label-switching phenomenon \citep{Stephens2000}. For this reason, one typically focuses on the so-called co-clustering matrix $\bm{S}$ \citep[e.g.][]{Fritsch2009}, whose entries $s_{ii'}$ are such that
\begin{equation*}
s_{ii'} = \mathds{P}(c_i = c_{i'} \mid \lambda, \bm{X}), \qquad i,i' \in  \{1,\dots,n\}.
\end{equation*}
Each probability $s_{ii'}$ can be easily approximated from \textsc{mcmc} samples and does not depend on the specific labels of $c_i$ and $c_{i'}$.  From the matrix $\bm{S}$, one can extract the probabilities
\begin{equation}\label{miscl_probabilities}
s_{ii^*_k} = \mathds{P}(c_i = c_{i_k^*} \mid \lambda, \bm{X}), \qquad i = 1,\dots,n, \quad k=1,\dots,K,
\end{equation}
with $i_k^*$ being the $k$th medoid. These probabilities can be viewed as surrogates for the unavailable $\mathds{P}(c_i = k \mid \lambda, \bm{X})$. The quantities $s_{ii^*_k}$ can be used to identify those units whose clustering allocation is uncertain, as in~Figure~\ref{fig:intro}. See  Section~\ref{sec:illustration} for further practical examples. 

Suppose a new data point $\bm{x}_{n+1}$ becomes available and we are interested in assigning it to a cluster. The allocation probabilities are readily available from Theorem~\ref{GibbsSampling}, and one has that
\begin{equation}\label{cluster_predictive}
\mathds{P}(c_{n+1} = k \mid  \lambda, \bm{X}, \bm{x}_{n+1}) \propto \frac{\rho(C_{k,n+1}; \lambda, \bm{X}_{k,n+1})}{\rho(C_k; \lambda, \bm{X}_k)}, \qquad k=1,\dots,K,
\end{equation}
where $C_{k,n+1}$ and $\bm{X}_{k,n+1}$ represent the $k$th index set and data matrix, respectively, having added the $n+1$th observation. 

A decision-theoretic approach for uncertainty quantification is described in \citet{Wade2018}, who propose point estimates and credible intervals for random partitions. Their ideas rely on the \emph{variation of information} (\textsc{vi}) distance $d_\textsc{vi}(\bm{c},\bm{c}')$ \citep{Meila2007}. To characterize uncertainty around a point estimate $\hat{\bm{c}}$ they consider a credible ball $\mathcal{B}_{1-\alpha}(\hat{\bm{c}})$ having probability $1 - \alpha$. Such a ball is constructed by adding partitions of increasing \textsc{vi} distance from $\hat{\bm{c}}$ until the posterior probability is greater than $1-\alpha$.  To represent the boundaries of $\mathcal{B}_{1-\alpha}(\hat{\bm{c}})$, \citet{Wade2018}
use \emph{horizontal bounds} $\hat{\bm{c}}_{1-\alpha}$ corresponding to the partitions in $\mathcal{B}_{1-\alpha}(\hat{\bm{c}})$ having the greatest \textsc{vi} distance from $\hat{\bm{c}}$.  If $\hat{\bm{c}}$ and 
$\hat{\bm{c}}_{1-\alpha}$ are similar, then posterior uncertainty in clustering is small.  

\section{Cohesions based on Bregman divergences}\label{sec:Bregman}

We now specialize the general \textsc{gb-ppm} formulation and describe a family of cohesion functions based on Bregman divergences. The $\hat{\bm{c}}_\textsc{map}$ of this \textsc{gb-ppm} can be obtained leveraging the Bregman k-means algorithm \citep{Banerjee2005}. This leads to very efficient computational routines for point estimation and enables uncertainty quantification for a wide class of existing methods, including the k-means example illustrated in Figure~\ref{fig:intro}. First let us recall the definition of Bregman divergence.

\begin{definition} Let $\varphi : \mathds{X} \rightarrow \mathds{R}$ be a strictly convex function defined on a convex set $\mathds{X} \subseteq \mathds{R}^d$, such that $\varphi$ is differentiable on the relative interior of $\mathds{X}$. A \emph{Bregman divergence} is defined as
\begin{equation*}
\mathcal{D}_\varphi(\bm{x}; \bm{\mu}) = \varphi(\bm{x}) - [\varphi(\bm{\mu}) + (\bm{x} - \bm{\mu})^\intercal \nabla\varphi(\bm{\mu})],
\end{equation*}
for any $\bm{x} \in \mathds{X}$ and any $\bm{\mu}$ in the relative interior of $\mathds{X}$, where $\nabla \varphi(\bm{\mu})$ represents the gradient vector of $\varphi$ evaluated at $\bm{\mu}$. 
\end{definition}

A Bregman divergence $\mathcal{D}_\varphi(\bm{x}; \bm{\mu})$ is always non-negative and it has a simple geometric interpretation: the ``distance'' between $\bm{x}$ and $\bm{\mu}$ is measured as the difference between $\varphi(\bm{x})$ and the value of its tangent hyperplane at $\bm{\mu}$, evaluated at~$\bm{x}$. Many well known discrepancies are special cases of Bregman divergences, including the squared Euclidean distance, the Mahalanobis distance, and the \textsc{kl}-divergence. In the following, we employ Bregman divergences to construct a \textsc{gb-ppm}.

\begin{definition}\label{Bregman} Let  $\pi_\varphi(\bm{c} \mid \lambda, \bm{X})$ be a \textsc{gb-ppm}. We will say it has \emph{Bregman cohesions} if
\begin{equation*}
\pi_\varphi(\bm{c} \mid \lambda, \bm{X}) \propto \prod_{k=1}^K \rho(C_k; \lambda, \bm{X}_k)  = \prod_{k=1}^K \exp\left\{ - \lambda \sum_{i \in C_k} \: \mathcal{D}_\varphi(\bm{x}_i; \bar{\bm{x}}_k)\right\}, \qquad \bm{c} : |\bm{C}| = K,
\end{equation*}
where $\mathcal{D}_\varphi(\bm{x}; \bm{\mu})$ is a Bregman divergence.
\end{definition}

Note that the arithmetic mean $\bar{\bm{x}}_k$ in the above definition is not an arbitrary choice, since it maximizes the associated cohesion function, that is, $\bar{\bm{x}}_k= \arg \max_{\bm{\mu}} \exp\{-\lambda \sum_{i \in C_k} \mathcal{D}_\varphi(\bm{x}_i; \bm{\mu})\}$. Moreover, recall from Proposition~\ref{map} that $\hat{\bm{c}}_\textsc{map}$ is obtained as the solution of the minimization problem
\begin{equation*}
\min_{\bm{c}\: : \: |\bm{C}| = K}\sum_{k=1}^K\sum_{i \in C_k} \mathcal{D}_\varphi(\bm{x}_i; \bar{\bm{x}}_k).
\end{equation*}
The above optimization can be solved through the efficient Bregman k-means algorithm, which is recalled in Algorithm~\ref{Algorithm}. Such a procedure can be regarded as a generalization of k-means and it entails essentially the same steps. Moreover, it can be shown that Algorithm~\ref{Algorithm} monotonically decreases the loss function, and it reaches a local optimum in a finite number of steps \citep{Banerjee2005}. Note also that in a \textsc{gb-ppm} with Bregman cohesion the ``generalized'' centroids $\bm{m}_1,\dots,\bm{m}_K$, defined in Section~\ref{sec:uq}, correspond to the within-cluster means and therefore are straightforward to compute. Indeed,  for any $k=1,\dots,K$ one has that $\bm{m}_k = \arg \min_{\bm{x} \in \mathds{X}}\mathcal{D}_\varphi(\bm{x}; \bar{\bm{x}}_k) = \bar{\bm{x}}_k$, because $\mathcal{D}_\varphi(\bm{x}; \bar{\bm{x}}_k) = 0$ if and only if $\bm{x} = \bar{\bm{x}}_k$. 

Finally, we remark that the Bregman divergence $\mathcal{D}_\varphi(\bm{x}_i; \bar{\bm{x}}_k)$ evaluated at $\bar{\bm{x}}_k$ is not always well-defined, because the arithmetic mean does not necessarily belong to the relative interior of $\mathds{X}$. This typically occurs when the data have discrete support, e.g. when $x_{ij} \in \{0,1\}$. This issue was not previously emphasized in \citet{Banerjee2005}, and it may lead to ill-behaved steps in Algorithm~\ref{Algorithm}. We will address this difficulty in Section~\ref{sec:boundary} by relying on adjusted centroids $\tilde{\bm{x}}_k$.

\begin{algorithm}[t]
\vspace{5pt}
Choose $K$ and a set of initial centroids $\bm{m}_1,\dots,\bm{m}_K$.\\ \vspace{5pt}
\do{Until the centroids stabilize:}{
\\
\vspace{5pt}
\For{$i=1,\dots,n$}{
	Set the cluster indicator $c_i$ equal to $k$, so that $\mathcal{D}_\varphi(\bm{x}_i; \bm{m}_k)$ is minimum.
\vspace{5pt}
}	
\For{$k=1,\dots,K$}{
	Let $\bm{m}_k$ be equal to the arithmetic mean $\bar{\bm{x}}_k$ of the subjects belonging to group $k$. 
\vspace{5pt}
}
}
\Return $\hat{\bm{c}}_\textsc{map} = (c_1, \ldots, c_n)$. \\
\caption{Bregman k-means \citep{Banerjee2005} \label{Algorithm}}
\end{algorithm}

\subsection{Connection with exponential dispersion families}\label{dispersion_families}

It turns out that the \textsc{gb-ppm} in Definition~\ref{Bregman} is deeply connected with mixtures of exponential dispersion models. Such a class of distributions, introduced by \citet{Jorgensen1987}, is a generalization of regular exponential families and is indexed by a \emph{dispersion parameter}. We show that $\lambda$ in Definition~\ref{Bregman} coincides with such a dispersion parameter. This alternative probabilistic representation helps in the interpretation of the loss and in the elicitation of $\lambda$, which is otherwise a difficult problem \citep{Bissiri2016,Holmes2017}. We refer to Section~\ref{sec:models} and Section~\ref{sec:illustration} for specific examples and default choices for $\lambda$. 
We remark that such a connection holds only for \emph{regular} Bregman divergences \citep{Banerjee2005}, which are, however, the vast majority of the known cases.

\begin{definition}\label{dispersion} Let $\pi(\bm{x} \mid \lambda)$  be a density function on $\mathds{X} \subseteq \mathds{R}^d$ indexed by $\lambda > 0$ and let $\Pi_\lambda$ be its probability measure. Then, the class of densities
\begin{equation*}
\pi_{\textsc{ed}}(\bm{x} \mid \bm{\theta},\lambda) = \pi(\bm{x} \mid \lambda) e^{\lambda [\bm{\theta}^\intercal\bm{x} - \kappa(\bm{\theta})]}, \qquad \bm{\theta} \in \Theta, \quad \lambda \in \Lambda,
\end{equation*}
is the \emph{exponential dispersion family}, where $\Theta \times \Lambda = \{ (\bm{\theta},\lambda) \in \mathds{R}^d \times \mathds{R}_+ \setminus \{0\} : \int_\mathds{X} e^{\lambda \bm{\theta}^\intercal\bm{x}}\Pi_\lambda(\dd \bm{x}) < \infty\}$.
\end{definition}

To clarify the role of the parameters $\lambda$ and $\bm{\theta}$, we recall the first two moments of an exponential dispersion random vector. In particular, if $\bm{x} \sim  \pi_{\textsc{ed}}(\bm{x} \mid \bm{\theta},\lambda)$, then
\begin{equation*}
\mathds{E}(\bm{x}) = \mu(\bm{\theta}), \qquad \text{Var}(\bm{x}) = \frac{1}{\lambda} \bm{V}(\bm{\theta}),
\end{equation*}
where the function $\mu(\cdot)$ is injective and $\bm{V}(\bm{\theta})$ is a $d \times d$ matrix not depending on $\lambda$. Hence, there is a one-to-one correspondence between the so-called \emph{natural parametrization} $\bm{\theta}$ and the \emph{mean parametrization} $\bm{\mu} = \mu(\bm{\theta})$, so that $\bm{\theta} = \theta(\bm{\mu}) = \mu^{-1}(\bm{\mu})$. Moreover, the parameters  $\lambda$ and $\bm{\theta}$ control the scale and the location of $\bm{x}$, respectively. The connection between exponential dispersion families and the \textsc{gb-ppm} with Bregman cohesion is explained in the next Theorem.

\begin{theorem}\label{profile_lik} Let $\pi_\textsc{ed}(\bm{c} \mid \lambda, \bm{X})$ be a \textsc{gb-ppm} of the form
\begin{equation*}
\begin{aligned}
\pi_\textsc{ed}(\bm{c} \mid \lambda, \bm{X}) &\propto \prod_{k=1}^K\prod_{i \in C_k} \pi_{\textsc{ed}}(\bm{x}_i \mid \hat{\bm{\theta}}_k, \lambda) \propto \prod_{k=1}^K \prod_{i \in C_k} \pi(\bm{x}_i \mid \lambda) \exp\left\{\lambda [\hat{\bm{\theta}}_k^\intercal\bm{x}_i - \kappa(\hat{\bm{\theta}}_k)]\right\}, \quad \bm{c} : |\bm{C}| = K,
\end{aligned}
\end{equation*}
where each $\hat{\bm{\theta}}_k = \theta(\bar{\bm{x}}_k) =\arg \max_{\bm{\theta}} \prod_{i \in C_k} \pi_{\textsc{ed}}(\bm{x}_i \mid \bm{\theta}, \lambda)$ is the maximum likelihood estimate of $\bm{\theta}_k$ for any given partition $\bm{c}$. Then, there exists a \textsc{gb-ppm} with Bregman cohesion as in Definition~\ref{Bregman} such that
\begin{equation*}
\pi_\textsc{ed}(\bm{c} \mid \lambda, \bm{X})  = \pi_\varphi(\bm{c} \mid \lambda, \bm{X}), \qquad \bm{c} : |\bm{C}| = K,
\end{equation*}
for some suitable function $\varphi$. 
\end{theorem}

Broadly speaking, any exponential dispersion family is associated to a \textsc{gb-ppm} Bregman cohesion. The Bregman divergences admitting such a representation are called \emph{regular}. The proof relies on the existence of a suitable function $\varphi(\bm{x})$ such that
\begin{equation*}
\lambda \bm{\theta}^\intercal\bm{x} - \lambda \kappa(\bm{\theta}) = - \lambda \mathcal{D}_\varphi(\bm{x}; \mu(\bm{\theta})) + \lambda \varphi(\bm{x}),
\end{equation*}
a result which has been formally proved by \citet{Banerjee2005}. 
Importantly, Theorem~\ref{profile_lik} also clarifies the role of $\lambda$, which is proportional to the within-cluster precision. In several practical cases $\lambda$ may be set equal to a known constant, but it can be also estimated from the data; see Section~\ref{sec:models} for examples and practical remarks. In addition, note that the maximum likelihood estimates $\hat{\bm{\theta}}_k = \theta(\bar{\bm{x}}_k)$ do not necessarily exist for all the partitions~$\bm{c}$. This indeed occurs whenever the associated Bregman divergence $\mathcal{D}_\varphi(\bm{x}; \bar{\bm{x}}_k)$ is not well-defined. These ill-behaved scenarios are discussed in Section~\ref{sec:boundary}.

Finally, a further consequence of Theorem~\ref{profile_lik} is that the \textsc{gb-ppm} $\pi_\varphi(\bm{c} \mid \lambda, \bm{X})$ can be regarded as the Bayesian update of a \emph{profile likelihood}, a likelihood function with nuisance parameters replaced by their maximum likelihood estimates. Thus, $\pi_\textsc{ed}(\bm{c} \mid \lambda, \bm{X})$ may be seen also as an ``approximate'' Bayesian model.  Moreover, note that $\hat{\bm{c}}_\textsc{map}$ can be actually interpreted as the maximum likelihood estimator. Although this connection may be helpful for the choice of $\lambda$,  we stress again that we rely on the justification in Section~\ref{sec:bissiri}, meaning that $\pi_\varphi(\bm{c} \mid \lambda, \bm{X})$ should be regarded as a proper update of beliefs rather than an approximate model.


\subsection{Discrete data}\label{sec:boundary}

Some difficulties can arise in Bregman clustering procedures if the centroid $\bar{\bm{x}}_k$ is outside the relative interior of $\mathds{X}$ or, equivalently, if the maximum likelihood estimate $\hat{\bm{\theta}}_k$ does not exist. Indeed, in such a case the Bregman divergence and the associated profile likelihood are not well-defined.  As previously mentioned, this issue is typically encountered whenever the data have discrete support.

When the sample size is large enough, most of the partitions are well-defined and therefore one could disregard the cluster solutions that lead to ill-defined losses. However, the representation of Theorem~\ref{profile_lik} suggests a more elegant fix. Specifically, each $\theta(\bar{\bm{x}}_k)$ can be replaced by a penalized maximum likelihood estimate $\tilde{\bm{\theta}}_k$, so that
\begin{equation*}
\tilde{\bm{\theta}}_k = \arg \max_{\bm{\theta}} \pi(\bm{\theta} \mid \lambda) \prod_{i \in C_k} \pi_{\textsc{ed}}(\bm{x}_i \mid \bm{\theta}, \lambda), \qquad k=1,\dots, K,
\end{equation*}
for a suitable penalty function $\pi(\bm{\theta} \mid \lambda)$ that leads to well-defined estimates. Equivalently, we can replace the arithmetic means $\bar{\bm{x}}_k$ with the adjusted centroids $\tilde{\bm{x}}_k = \mu(\tilde{\bm{\theta}}_k)$, so that $\tilde{\bm{\theta}}_k = \theta(\tilde{\bm{x}}_k)$. To preserve the \textsc{gb-ppm} structure, we require that the estimates $\tilde{\bm{\theta}}_k$ do not depend on $\lambda$. 

In exponential dispersion families, a convenient choice for the penalty function $\pi(\bm{\theta} \mid \lambda)$ is given by \citet{Firth1993}. In such a case, the estimates $\theta(\tilde{\bm{x}}_k)$ have improved asymptotic properties and are typically easy to compute. Moreover, the penalty function $\pi(\bm{\theta} \mid \lambda)$ can be interpreted as the Jeffrey's prior associated to $\pi_{\textsc{ed}}(\bm{x} \mid \bm{\theta}, \lambda)$. Since the adjusted centroids $\tilde{\bm{x}}_k$ do not depend on $\lambda$, we then obtain the following \textsc{gb-ppm} 
\begin{equation}\label{adjusted_gbppm}
\pi_\varphi(\bm{c} \mid \lambda, \bm{X}) \propto \prod_{k=1}^K \prod_{i \in C_k} \pi_{\textsc{ed}}(\bm{x}_i \mid \theta(\tilde{\bm{x}}_k), \lambda) \propto \prod_{k=1}^K \exp\left\{ - \lambda \sum_{i \in C_k} \: \mathcal{D}_\varphi(\bm{x}_i; \tilde{\bm{x}}_k)\right\}, \qquad \bm{c} : |\bm{C}| = K,
\end{equation}
which is well-defined for any choice of the partition $\bm{c}$.  Importantly, the $\hat{\bm{c}}_\textsc{map}$ solution can be still obtained using Algorithm~\ref{Algorithm3}, which is a simple modification of Bregman k-means that preserves its main properties. This is clarified in the following Proposition.

\begin{algorithm}[t]
\vspace{5pt}
Choose $K$ and a set of initial centroids $\bm{m}_1,\dots,\bm{m}_K$.\\ \vspace{5pt}
\do{Until the centroids stabilize:}{
\\
\vspace{5pt}
\For{$i=1,\dots,n$}{
	Set the cluster indicator $c_i$ equal to $k$, so that $\mathcal{D}_\varphi(\bm{x}_i; \bm{m}_k)$ is minimum.
\vspace{5pt}
}	
\For{$k=1,\dots,K$}{
	Let $\bm{m}_k$ be equal to the adjusted estimate $\tilde{\bm{x}}_k$ of the subjects belonging to group $k$. 
\vspace{5pt}
}
}
\Return $\hat{\bm{c}}_\textsc{map} = (c_1, \ldots, c_n)$. \\
\caption{Adjusted Bregman k-means \label{Algorithm3}}
\end{algorithm}

\begin{proposition}\label{monotone}
The Bregman k-means with adjusted centroids in Algorithm~\ref{Algorithm3} monotonically decreases the loss function $\ell(\bm{c};\bm{X}) = \sum_{k=1}^K\sum_{i \in C_k} \mathcal{D}_\varphi(\bm{x}_i; \tilde{\bm{x}}_k)$ for any $\bm{c} : |\bm{C}| =K$. Moreover, Algorithm~\ref{Algorithm3} terminates in a finite number of steps at a partition that is locally optimal. 
\end{proposition}

\section{Cohesions based on pairwise dissimilarities}\label{sec:distance}

We now specialize the general \textsc{gb-ppm} formulation in a different direction, and we describe a family of cohesion functions based on pairwise dissimilarities. Such a family may offer a more robust clustering framework compared to Bregman cohesions. The robustness properties of this \textsc{gb-ppm} will depend on the chosen dissimilarity measure. Within the algorithmic framework, the closest relative to this \textsc{gb-ppm} approach is the so-called k-medoids or \textsc{pam} algorithm \citep{Kaufman1990,Hastie2008}, which indeed may be used to mitigate the drawbacks of k-means. As we shall see, this class is also closely related to the approximate Bayesian model of \citet{Duan2019}.

Let us assume that the covariate space is $\mathds{X} = \mathds{R}^d$ and let $||\bm{x}||_p = (|x_1|^p + \cdots + |x_d|^p)^{1/p}$ be the $L^p$ norm, for any $p \ge 1$ and $\bm{x} \in \mathds{R}^d$. Then, a general measure of dissimilarity is
\begin{equation*}
\gamma(||\bm{x}_i - \bm{x}_{i'}||_p^p), \qquad \bm{x}_i, \bm{x}_{i'} \in \mathds{R}^d,
\end{equation*} 
for some increasing function $\gamma : \mathds{R}_+ \rightarrow \mathds{R}_+$ such that $\gamma(0) = 0$. In general $\gamma(||\bm{x}_i - \bm{x}_{i'}||_p^p)$ is not a metric on $\mathds{R}^d$, although this could be the case for specific choices of $\gamma$. For example, with $\gamma(x) = x^{1/p}$ one obtains the Minkowski distance. The squared Euclidean distance is recovered when $\gamma(x) = x$ and $p = 2$. Recalling the definition of a \textsc{gb-ppm}, we then seek a discrepancy function measuring the distance of the $i$th unit from the $k$th cluster. We consider the so-called average dissimilarity
\begin{equation}\label{average}
\mathcal{D}_\gamma(\bm{x}_i; \bm{X}_k) =  \frac{1}{n_k} \sum_{i' \in C_k} \gamma(||\bm{x}_i - \bm{x}_{i'}||_p^p),
\end{equation}
which leads to the following novel \textsc{gb-ppm}.  

\begin{definition}\label{pairwise} Let  $\pi_\gamma(\bm{c} \mid \lambda, \bm{X})$ be a \textsc{gb-ppm} with covariate space $\mathds{X} = \mathds{R}^d$. We will say it has \emph{average dissimilarity cohesions} if
\begin{equation*}
\pi_\gamma(\bm{c} \mid \lambda, \bm{X}) \propto \prod_{k=1}^K \rho(C_k; \lambda, \bm{X}_k)  = \prod_{k=1}^K \exp\left\{ - \frac{\lambda}{2}\sum_{i \in C_k} \frac{1}{n_k}\sum_{i' \in C_k} \gamma(||\bm{x}_i - \bm{x}_{i'}||_p^p)\right\}, \qquad \bm{c} : |\bm{C}| = K,
\end{equation*}
with $p \ge 1$ and $\gamma : \mathds{R}_+ \rightarrow \mathds{R}_+$ an increasing function such that $\gamma(0) = 0$. 
\end{definition}


\begin{algorithm}[t]
\vspace{5pt}
Randomly allocate the indicators $c_1,\dots,c_n$ into $K$ sets.\\ \vspace{5pt}
\do{Until the partition stabilizes:}{
\\
\vspace{5pt}
\For{$i=1,\dots,n$}{
	Allocate the indicator $c_i$, given the others $\bm{c}_{-i}$, to the $k$ cluster, so that
	$$\sum_{i' \in C_k} \mathcal{D}_\gamma(\bm{x}_{i'}; \bm{X}_k) - \sum_{i' \in C_{k,-i}} \mathcal{D}_\gamma(\bm{x}_{i'}; \bm{X}_{k,-i})$$ 
	is minimum. This difference can be computed efficiently exploiting recursive formulas; see Appendix~\ref{app1} for details.  
\vspace{5pt}
}
}
\Return $\hat{\bm{c}}_\textsc{map} = (c_1, \ldots, c_n)$. \\
\caption{k-dissimilarities \label{Algorithm4}}
\end{algorithm}

Posterior inference in a \textsc{gb-ppm} with average dissimilarity cohesions is not as fast as in the Bregman divergence case, but it is still quite efficient. Indeed, one needs to evaluate all the distinct $n(n-1)/2$ pairwise dissimilarities, which could be a computational bottleneck. However, if these values are pre-computed and stored, then inference can be easily conducted. Gibbs sampling is performed by iteratively drawing values from the full-conditionals in Theorem~\ref{GibbsSampling}. Moreover, an efficient routine for finding $\hat{\bm{c}}_\textsc{map}$ is available and is described in Algorithm~\ref{Algorithm4}. We call this novel method \emph{k-dissimilarities}, which might be of independent interest. Such an algorithm retains the main properties of the k-means algorithm, as clarified in the following Proposition. 

\begin{proposition}\label{monotone2}
The iterative allocations in the k-dissimilarities Algorithm~\ref{Algorithm4} monotonically decrease the loss function $\ell(\bm{c};\bm{X}) = \sum_{k=1}^K\sum_{i \in C_k} \mathcal{D}_\gamma(\bm{x}_i; \bm{X}_k)$ for any $\bm{c} : |\bm{C}| =K$. Moreover, Algorithm~\ref{Algorithm4} terminates in a finite number of steps at a partition that is locally optimal. 
\end{proposition}

The results of Proposition~\ref{monotone2} are not surprising, because at each step the k-dissimilarities algorithm locally minimizes the loss function. Moreover, note that the discrepancy $\sum_{i \in C_k} \mathcal{D}_\gamma(\bm{x}_i; \bm{X}_k)$ should not be re-computed at each step, thanks to a recursive formula. Such a recursion, described in Appendix~\ref{app1}, is extremely useful also in the Gibbs sampling implied by Theorem~\ref{GibbsSampling}.

\subsection{Connection with spherical distributions}\label{composite_likelihoods}

It turns out that \textsc{gb-ppm}s with pairwise dissimilarities are deeply connected with $L^p$ spherical distributions \citep{Gupta1997}. Such a class of random vectors in $\mathds{R}^d$ is a generalization of spherical distributions in~$L^2$; see \citet{Fang1990} for an overview. Paralleling the discussion of Section~\ref{dispersion_families}, these probabilistic connections improve the interpretability of the model and facilitate the choice or the estimation of $\lambda$. This will be discussed in detail in Section~\ref{sec:models}. We begin our discussion by providing the definition of $L^p$ spherical distributions. 

\begin{definition} A random vector $\bm{x} \in \mathds{R}^d$ follows a $L^p$ spherical distribution if its density function can be written as $\pi_\textsc{sp}(\bm{x}) = g(||\bm{x}||_p^p)$ for some measurable function $g : \mathds{R}_+ \rightarrow \mathds{R}_+$. 
\end{definition}

The class of $L^p$ spherical distributions is very general as it includes the multivariate Gaussian, the multivariate Laplace, and the multivariate Student's t, among others. Such a family is indexed by the function $g$, which is sometimes called the \emph{density generator}. The connection with the \textsc{gb-ppm} of Definition~\ref{pairwise} is clarified in the following Theorem.

\begin{theorem}\label{pairwise_lik} Let $\pi_\gamma(\bm{c} \mid \lambda, \bm{X})$ be a \textsc{gb-ppm} with average dissimilarities as in Definition~\ref{pairwise}. If
\begin{equation*}
\int_{\mathds{R}_+} r^{d-1} \exp\left\{-\frac{\lambda}{2} \gamma(r^p)\right\}\dd r < \infty,
\end{equation*}
then there exists an $L^p$ spherical distribution on $\mathds{R}^d$ such that
\begin{equation*}
\pi_\gamma(\bm{c} \mid \lambda, \bm{X}) \propto \prod_{k=1}^K \prod_{i \in C_k} \left[\prod_{i' \in C_k} \pi_\textsc{sp}(\bm{x}_i - \bm{x}_{i'} \mid \lambda)\right]^{1/n_k},
\end{equation*}
where $\pi_\textsc{sp}(\bm{x}_i - \bm{x}_{i'} \mid \lambda) \propto \exp\left\{-\lambda/2\gamma(||\bm{x}_i - \bm{x}_{i'}||_p^p)\right\}$ for any $i\in C_k$ and $i' \in C_k$. 
\end{theorem}

Hence, if a simple integrability condition on the $\gamma$ function holds true, then the \textsc{gb-ppm} of Definition~\ref{pairwise} has a simple probabilistic interpretation. In addition, Theorem~\ref{pairwise_lik} implies that $\exp\{-\lambda\gamma(x)\}$ is proportional to a density generator for $L^p$ spherical distributions. The role of $\lambda$ will depend on the specific choices of $\gamma$, but we will show in Section~\ref{sec:models} that $\lambda$ can often be interpreted as a scale parameter.

Importantly, Theorem~\ref{pairwise_lik} highlights that a \textsc{gb-ppm} with average dissimilarities can be interpreted as the Bayesian update of a \emph{pairwise difference likelihood} \citep{Lele2002,Varin2011}, a special instance of composite likelihood. Suppose that, conditionally on the cluster indicators, the observations follow some location family of distributions, namely
\begin{equation*}
(\bm{x}_i \mid \bm{\mu}_k, \lambda, c_i = k) \overset{\textup{iid}}{\sim} \pi\left(\bm{x} - \bm{\mu}_k \mid \lambda \right), \qquad i \in C_k, \quad k=1,\dots,K,
\end{equation*}
where $\bm{\mu}_k \in \mathds{R}^d$. Following \citet{Lele2002,Varin2011} one could consider a within-cluster pairwise difference likelihood, which is based on the differences $\bm{x}_i - \bm{x}_{i'}$, for any $i \in C_k$ and $i' \in C_k$. Note that the within-cluster differences $\bm{x}_i - \bm{x}_{i'}$ are identically distributed and they do not depend on the location parameter $\bm{\mu}_k$.  $L^p$ spherical distributions are indeed an appealing and very general modeling choice for $\bm{x}_i - \bm{x}_{i'}$, e.g. because they are symmetric around $0$, a natural requirement in this setting. Hence, the associated pairwise difference likelihood is proportional to
\begin{equation*}
\pi_\textsc{diff}(\bm{X} \mid \bm{c}, \lambda) \propto \prod_{k=1}^K \prod_{i \in C_k} \left[\prod_{i' \in C_k}\pi_\textsc{sp}(\bm{x}_i - \bm{x}_{i'} \mid \lambda)\right]^{1/n_k},
\end{equation*}
where the exponent $1/n_k$ is a correction that deflates the likelihood so that it has the usual asymptotic order. The key advantage of this kind of composite likelihood is that the location parameters have been removed.  As may already be clear, the \textsc{gb-ppm} in Definition~\ref{pairwise} corresponds to the Bayesian update of $\pi_\textsc{diff}(\bm{X} \mid \bm{c}, \lambda)$ if a suitable integrability condition holds. As discussed before, note that  we are not regarding $\pi_\gamma(\bm{c} \mid \lambda, \bm{X})$ as an approximate Bayesian model based on composite likelihoods as in \citet{Pauli2011} or \citet{Aliverti2020}. Instead, we stress once again that we rely on the coherent update of beliefs presented in Section~\ref{sec:bissiri}.

\section{Modeling examples}\label{sec:models}

In this section we present specific \textsc{gb-ppm}s belonging to the classes discussed in Sections~\ref{sec:Bregman} and \ref{sec:distance}. In doing so, we also address the issue of either specifying or estimating the value of $\lambda$. We suggest default strategies based on the theoretical findings of the previous sections. We emphasize that the models presented here were chosen for the sake of illustration. Indeed, we think that a core contribution of our approach is the generality of \textsc{gb-ppm}s, and therefore we are not claiming the superiority of these specific cases over alternative generalized Bayes models. Instead, we aim at providing some guidelines for their construction, usage and interpretation.

\subsection{Squared euclidean distance}\label{sec:squared_euclidean}

The first \textsc{gb-ppm} we describe relies on the quadratic loss given in equation~\eqref{kmeans_loss}, which is closely related to k-means. Given the central role the k-means algorithm plays in clustering problems, we believe the associated \textsc{gb-ppm} deserves special attention. If $\mathds{X} = \mathds{R}^d$, then it is easy to show that the loss~\eqref{kmeans_loss} is a Bregman divergence, implying that the associated \textsc{gb-ppm} is
\begin{equation*}\label{kmeans_gbppm}
\pi_\varphi(\bm{c} \mid \lambda, \bm{X}) \propto \prod_{k=1}^K  \exp\left\{- \lambda \sum_{i \in C_k} ||\bm{x}_i - \bar{\bm{x}}_k||_2^2\right\}, \qquad \bm{c} : |\bm{C}| = K,
\end{equation*}
with $\varphi(\bm{x}) = ||\bm{x}||_2^2$. The above model can also be regarded as a \textsc{gb-ppm} with pairwise dissimilarities. Recall  the well known identity $\lambda \sum_{i \in C_k} ||\bm{x}_i - \bar{\bm{x}}_k||_2^2 = \lambda/2  \sum_{i \in C_k}   n_k^{-1} \sum_{i' \in C_k} ||\bm{x}_i - \bm{x}_{i'}||_2^2$. Then, with $\gamma(x) = x$ and $p=2$ we can equivalently let
\begin{equation*}
\pi_\gamma(\bm{c} \mid \lambda, \bm{X}) \propto \prod_{k=1}^K \exp\left\{ - \frac{\lambda}{2}\sum_{i \in C_k} \frac{1}{n_k}\sum_{i' \in C_k} ||\bm{x}_i - \bm{x}_{i'}||_2^2\right\}, \qquad \bm{c} : |\bm{C}| = K.
\end{equation*}
Hence, the point estimate $\hat{\bm{c}}_\textsc{map}$ can be either obtained through Algorithm~\ref{Algorithm} or~\ref{Algorithm4}. The former coincides with the k-means algorithm of \citet{Lloyd1982}, whereas the latter is essentially a variation of the algorithm of \citet{Hartigan1979}.

The k-means \textsc{gb-ppm} therefore inherits the probabilistic interpretation of Theorems~\ref{profile_lik} and~\ref{pairwise_lik}. Specifically, the associated profile and composite likelihoods are consistent with the generative mechanism
\begin{equation}\label{kmeans_profile}
(\bm{x}_i \mid \bm{\mu}_k, \lambda, c_i = k) \overset{\textup{iid}}{\sim} \mbox{N}(\bm{\mu}_k, (2\lambda)^{-1} I_d), \qquad i \in C_k, \quad k=1,\dots,K,
\end{equation}
where $\mbox{N}(\bm{\mu},\bm{\Sigma})$ denotes a $d$-dimensional multivariate Gaussian with mean $\bm{\mu}$ and covariance matrix~$\bm{\Sigma}$. The multivariate Gaussian distribution in~\eqref{kmeans_profile} is indeed an exponential dispersion family. In addition, it follows that the $L^2$ spherical distribution characterizing the pairwise distances $\bm{x}_i - \bm{x}_{i'}$ must be a multivariate Gaussian, namely
\begin{equation}\label{kmeans_diff}
(\bm{x}_i - \bm{x}_{i'} \mid \lambda, c_i = k, c_{i'} = k) \sim \mbox{N}\left(\bm{0}, \lambda^{-1} I_d\right),
\end{equation}
for any $i \in C_k$, $i' \in C_k$ and $k=1,\dots,K$. Note that the profile likelihood obtained from equation~\eqref{kmeans_profile} coincides with the pairwise difference likelihood based on equation~\eqref{kmeans_diff}. 

Exploiting these probabilistic connections, we discuss a reasonable default strategy for the estimation of $\lambda$. Following \citet{Bissiri2016}, we incorporate $\lambda$ in the loss function and we specify a hierarchical prior $\pi(\lambda)$. The balance between the joint prior $\pi(\bm{c})\pi(\lambda)$ and the joint loss $\ell(\bm{c},\lambda;\bm{X})$ is regulated by a parameter $\tilde{\lambda} > 0$, whose choice is simpler than the one for $\lambda$. More precisely, we consider the following joint Gibbs posterior 
\begin{equation}\label{gibbs_joint}
\begin{aligned}
\pi(\bm{c}, \lambda \mid \tilde{\lambda}, \bm{X}) &\propto \pi(\bm{c}) \pi(\lambda) \exp\{- \tilde{\lambda} \ell(\bm{c},\lambda;\bm{X}) \}, \quad \text{with} \quad \ell(\bm{c},\lambda;\bm{X}) = \lambda\ell(\bm{c};\bm{X}) - \xi \log(\lambda), \\
\end{aligned}
\end{equation}
with $\bm{c} : |\bm{C}| = K$ and for some $\xi \ge 0$.  Theorems~\ref{profile_lik} and~\ref{pairwise_lik} suggest that joint inference about $\bm{c}$ and $\lambda$ should be based on the prior updated with the appropriate pseudo-likelihood as 
\begin{equation}\label{kmeans_joint}
\pi(\bm{c}, \lambda \mid \bm{X}) \propto \pi(\lambda)  \lambda^{nd/2}  \prod_{k=1}^K \exp\left\{- \lambda \sum_{i \in C_k} ||\bm{x}_i - \bar{\bm{x}}_k||_2^2\right\},\qquad \bm{c} : |\bm{C}| = K.
\end{equation}
Comparing equation~\eqref{gibbs_joint} with \eqref{kmeans_joint} we see that $\xi = nd/2$ and $\tilde{\lambda} = 1$ are natural default choices. Note that the additional term $\lambda^{nd/2}$ follows from Theorems~\ref{profile_lik} and~\ref{pairwise_lik}. Moreover, if we let $\lambda \sim \textsc{gamma}(a_\lambda, b_\lambda)$ a priori, then the full conditional $\pi(\lambda \mid \bm{c}, \bm{X})$ is still a Gamma with updated parameters $a_\lambda + nd/2$ and $b_\lambda + \sum_{k=1}^K \sum_{i \in C_k} ||\bm{x}_i - \bar{\bm{x}}_k||_2^2$, leading to a simple Gibbs sampling step. 

We remark that the choices $\xi = nd/2$ and $\tilde{\lambda} = 1$ represent a reasonable default, but alternatives might be considered. For example, in the same spirit of \citet{Miller2019}, one could let $\xi = nd/2$ and $0 < \tilde{\lambda} < 1$, thus obtaining a variant of the joint model~\eqref{kmeans_joint} which deflates the importance of the pseudo-likelihood. The parameter $\tilde{\lambda}$ can be then selected either subjectively or following the guidelines presented in their paper.

\subsection{Minkowski distance}\label{sec:Minkowski}

As a more robust alternative to the k-means case, we present a 
 \textsc{gb-ppm} based on pairwise dissimilarities, which essentially 
 replaces the squared Euclidean metric with the $L^p$ distance. For $p=1$ this makes the clustering results less sensitive to outliers.
 More precisely, if $\gamma(x) = x^{1/p}$ then $\gamma(||\bm{x}_i - \bm{x}_{i'}||_p^p) = ||\bm{x}_i - \bm{x}_{i'}||_p$ is the Minkowski distance, which is also called Manhattan distance when $p=1$. Then,
\begin{equation*}
\pi_\gamma(\bm{c} \mid \lambda, \bm{X}) \propto \prod_{k=1}^K \exp\left\{ - \frac{\lambda}{2}\sum_{i \in C_k} \frac{1}{n_k}\sum_{i' \in C_k} ||\bm{x}_i - \bm{x}_{i'}||_p\right\}, \qquad \bm{c} : |\bm{C}| = K.
\end{equation*}
This specific \textsc{gb-ppm} is closely related to a special case of \citet{Duan2019}, the only differences being the prior $\pi(\bm{c})$ and the condition $\bm{c} : |\bm{C}| = K$. However, we remark that we regard $\pi_\gamma(\bm{c} \mid \lambda, \bm{X})$  as a generalized Bayes posterior rather than a ``partially specified'' Bayesian model. Within our framework, the derivation of $\pi_\gamma(\bm{c} \mid \lambda, \bm{X})$ is very natural, being the combination of the factorized loss~\eqref{general_loss} and the arithmetic average given in~\eqref{average}. Hence, when $\gamma(x) = x^{1/p}$ our contribution leads to an alternative derivation of the model of \citet{Duan2019}, to the development of novel algorithms, and to an important probabilistic interpretation based on composite likelihoods. 

 It is easy to check that the condition $\int_{\mathds{R}_+} r^{d-1} e^{-\lambda/2 r}\dd r < \infty$ required by Theorem~\ref{pairwise_lik} is satisfied. Moreover, the $L^p$ spherical distribution associated to each pairwise difference $\bm{x}_i - \bm{x}_{i'}$, for any $i \in C_k, i' \in C_k$, has density
\begin{equation*}
\pi_\textsc{sp}(\bm{x}_i - \bm{x}_{i'} \mid \lambda) = \frac{p^{d-1} }{2^d \Gamma(1/p)^d} \frac{\Gamma(d/p)}{\Gamma(d)}\left(\frac{\lambda}{2}\right)^d \exp\left\{ - \frac{\lambda}{2}||\bm{x}_i - \bm{x}_{i'}||_p \right\}.
\end{equation*}
Thus, $\lambda$ can be interpreted as a scale parameter. The normalizing constant of the above density is related to the integral $\int_{\mathds{R}_+} r^{d-1} e^{-\lambda/2 r}\dd r$ \citep{Gupta1997}, which has an explicit solution.  Moreover, for any $p > 1$, the components of the above random vector are dependent, but when $p = 1$ we recover the law of $d$ independent Laplace distributions. Finally, simple properties of $L^p$ spherical distributions imply that $||\bm{x}_i - \bm{x}_{i'}||_p \sim \textsc{gamma}(d, \lambda/2)$ for any $p \ge 1$. 

We again suggest to estimate $\lambda$ from the data. Specifically, joint inference on $\lambda$ and $\bm{c}$ can be conducted as in equation~\eqref{gibbs_joint}, i.e. through a hierarchical approach. On the other hand, Theorem~\ref{pairwise_lik} suggests that joint inference should be based on the update of
\begin{equation*}
\pi(\bm{c}, \lambda \mid \bm{X}) \propto \pi(\lambda) \lambda^{n d} \prod_{k=1}^K \exp\left\{ - \frac{\lambda}{2}\sum_{i \in C_k} \frac{1}{n_k}\sum_{i' \in C_k} ||\bm{x}_i - \bm{x}_{i'}||_p\right\}, \qquad \bm{c} : |\bm{C}| = K,
\end{equation*}
where the additional term $\lambda^{nd}$ follows from the pairwise difference likelihood representation. This leads to the default values $\xi = n d$ and $\tilde{\lambda} = 1$. The prior $\lambda \sim \textsc{gamma}(a_\lambda, b_\lambda)$ is a computationally convenient choice also in this case, since it leads to a conjugate Gibbs sampling step.

\subsection{Kullback-Leibler divergence}\label{sec:kl_model}
The last \textsc{gb-ppm} model we describe involves discrete data. Specifically, suppose $\mathds{X} = \{0,1\}^d$, meaning that $\bm{x}_i =(x_{i1},\dots,x_{id})^\intercal$ is a collection of binary indicators. The nature of the data leads to ill-behaved losses, as discussed in Section~\ref{sec:boundary}. To overcome these difficulties, we rely on a \textsc{gb-ppm} with adjusted centroids $\tilde{\bm{x}}_k = (\tilde{x}_{k1},\dots,\tilde{x}_{kd})^\intercal$, as in equation~\eqref{adjusted_gbppm}, and we let
\begin{equation}\label{kl_gbppm}
\begin{aligned}
\pi_\varphi(\bm{c} \mid \lambda=1, \bm{X}) &\propto  \prod_{k=1}^K \exp\left\{- \sum_{i \in C_k}\sum_{j=1}^d \textsc{kl}(p_{ij} \mid\mid \tilde{p}_{kj})\right\}, \quad \bm{c} :|\bm{C}| = K,\\
 &\propto \prod_{k=1}^K  \prod_{i \in C_k} \exp\left\{\sum_{j=1}^d\left[ x_{ij} \log\left(\frac{\tilde{x}_{kj}}{1 - \tilde{x}_{kj}}\right) + \log(1 - \tilde{x}_{kj})\right]\right\}, \quad \bm{c} :|\bm{C}| = K, 
\end{aligned}
\end{equation}
where each $p_{ij} = (x_{ij},1- x_{ij})^\intercal$ and $\tilde{p}_{kj} = (\tilde{x}_{kj},1-\tilde{x}_{kj})^\intercal$ are vectors of  probabilities. The Kullback-Leibler divergence is well-defined even if $p_{ij}$ are degenerate probabilities. The dispersion parameter $\lambda = 1$ is fixed. Considering equation~\eqref{kl_gbppm}, the 
kernel $\pi_{\textsc{ed}}(\bm{x} \mid \bm{\theta},\lambda=1)$ is that of $d$ independent Bernoulli random variables, linking this \textsc{gb-ppm} with latent class models \citep{Lazarsfeld1968}. The adjusted centroids are obtained exploiting the \citet{Firth1993} correction for  $\theta(\tilde{\bm{x}}_k)$, which leads to
\begin{equation*}
\tilde{x}_{kj} = \frac{n_k}{n_k +1}\bar{x}_{kj} + \frac{1}{2}\frac{1}{n_k +1}, \qquad j=1,\dots,d, \quad k=1,\dots,K, 
\end{equation*}
where $\bar{x}_{kj}$ denotes the within-cluster proportion. Boundary issues are avoided because $\tilde{x}_{kj} \in (0,1)$.  Equation~\eqref{kl_gbppm} relies on the pseudo-likelihood representation from Theorem~\ref{profile_lik}, which implies that $\lambda = 1$.

\section{Illustrations}\label{sec:illustration}

In this section, we further illustrate \textsc{gb-ppm}s through synthetic and real data examples. For simplicity, we focus on the models described in Section~\ref{sec:models}. In Section~\ref{sec:sim1} we illustrate the uncertainty quantification tools described in Section~\ref{sec:uq} on a synthetic dataset and compare the results with ``oracle'' clustering probabilities associated with the true generative mechanism. In Section~\ref{sec:sim2} we show that \textsc{gb-ppm}s with pairwise dissimilarities improve on robustness compared with k-means based approaches.
In Section~\ref{sec:realdata} we illustrate the performance of a \textsc{gb-ppm} model for binary data with a medical application  considered in \citet[][Chap. 13]{Agresti2002}. Moreover, we show that alternative point estimates, such as those presented in \citet{Wade2018}, can be more appropriate than the \textsc{map}.

\subsection{Uncertainty quantification and oracle clustering}\label{sec:sim1}

In this first simulation study we focus on the \textsc{gb-ppm} with squared Euclidean loss described in Section~\ref{sec:squared_euclidean}.  We consider $n = 200$ observations evenly divided in $K = 4$ clusters, each having $n_1 = \cdots = n_4 = 50$ data points. Within each partition we let
\begin{equation}\label{sim1}
(\bm{x}_i \mid \bm{\mu}_k, \sigma^2, c_i = k) \overset{\textup{iid}}{\sim} \mbox{N}\left(\bm{\mu}_k, \sigma^2 I_2\right), \qquad i \in C_k, \quad k=1,\dots,K,
\end{equation}
with $\bm{\mu}_1 = (-2,-2)$, $\bm{\mu}_2 = (-2,2)$, $\bm{\mu}_3 = (2,-2)$, and $\bm{\mu}_4 = (2,2)$. The within-cluster variance $\sigma^2$ takes the values reported in Table~\ref{tab1}. We let $K = 4$ for simplicity and we use the default strategy described in Section~\ref{sec:squared_euclidean}, with $\xi = nd/2$, $\tilde{\lambda}=1$, and $\pi(\lambda) \propto 1$.  Points estimates are obtained using k-means, and for uncertainty quantification we run the Gibbs sampler of Theorem~\ref{GibbsSampling} starting at the k-means estimate. We obtained $5,000$ posterior samples, after a burn-in of $1,000$ iterations. 
The traceplot of the loss function displays excellent mixing and no evidence against convergence. Horizontal bounds were computed through the \texttt{mcclust.ext} R package. 

\begin{figure}[t]
\includegraphics[width=\textwidth]{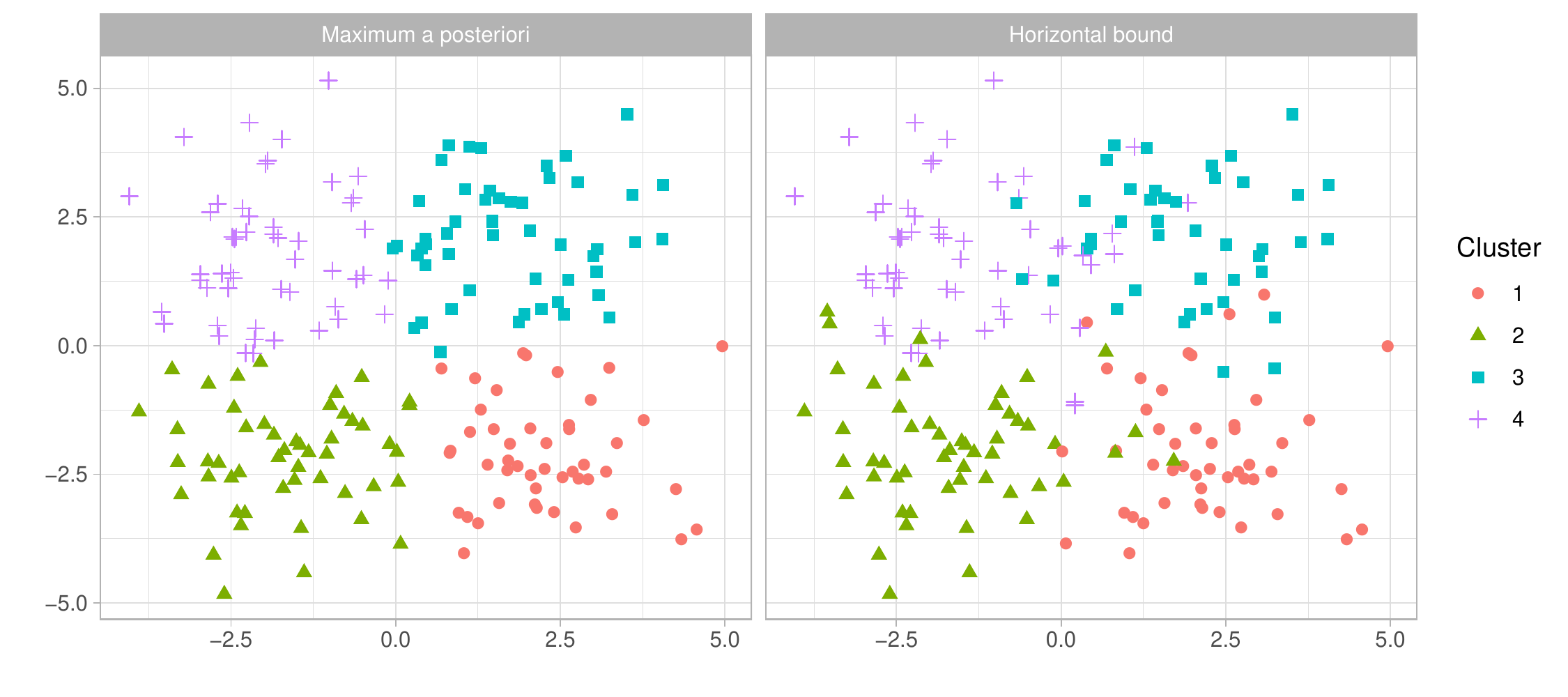}
\caption{Synthetic data points generated from equation~\eqref{sim1} with $\sigma^2 = 1.5$. Left plot: colors and shapes represent different clusters in the k-means point estimate $\hat{\bm{c}}_\textsc{map}$. Right plot: colors and shapes represent different clusters in the horizontal bound $\hat{\bm{c}}_{1-\alpha,\textsc{map}}$, with $\alpha = 0.05$. \label{fig:sim1_hb}}
\end{figure}

We display in Figure~\ref{fig:sim1_hb} the generated data points when $\sigma^2 = 1.5$, together with the k-means estimate $\hat{\bm{c}}_\textsc{map}$ and its horizontal bound $\hat{\bm{c}}_{1-\alpha,\textsc{map}}$, with $\alpha = 0.05$.  Recall that 
 $\hat{\bm{c}}_{1-\alpha,\textsc{map}}$ provides the clustering that is furthest from the k-means solution $\hat{\bm{c}}_\textsc{map}$ while still falling within a 95\% credible region around $\hat{\bm{c}}_\textsc{map}$.   Figure~\ref{fig:sim1_hb} shows that $\hat{\bm{c}}_{1-\alpha,\textsc{map}}$ is roughly similar to $\hat{\bm{c}}_\textsc{map}$ but many of the points near the cluster 
boundaries are assigned to different groups.
As evidenced in Table~\ref{tab1}, the variation of information between the \textsc{map} and its horizontal bound depends on the variability of the data, as expected. In particular, more dispersed observations are associated with wider credible intervals, and vice versa.

\begin{figure}[p]
\includegraphics[width=\textwidth]{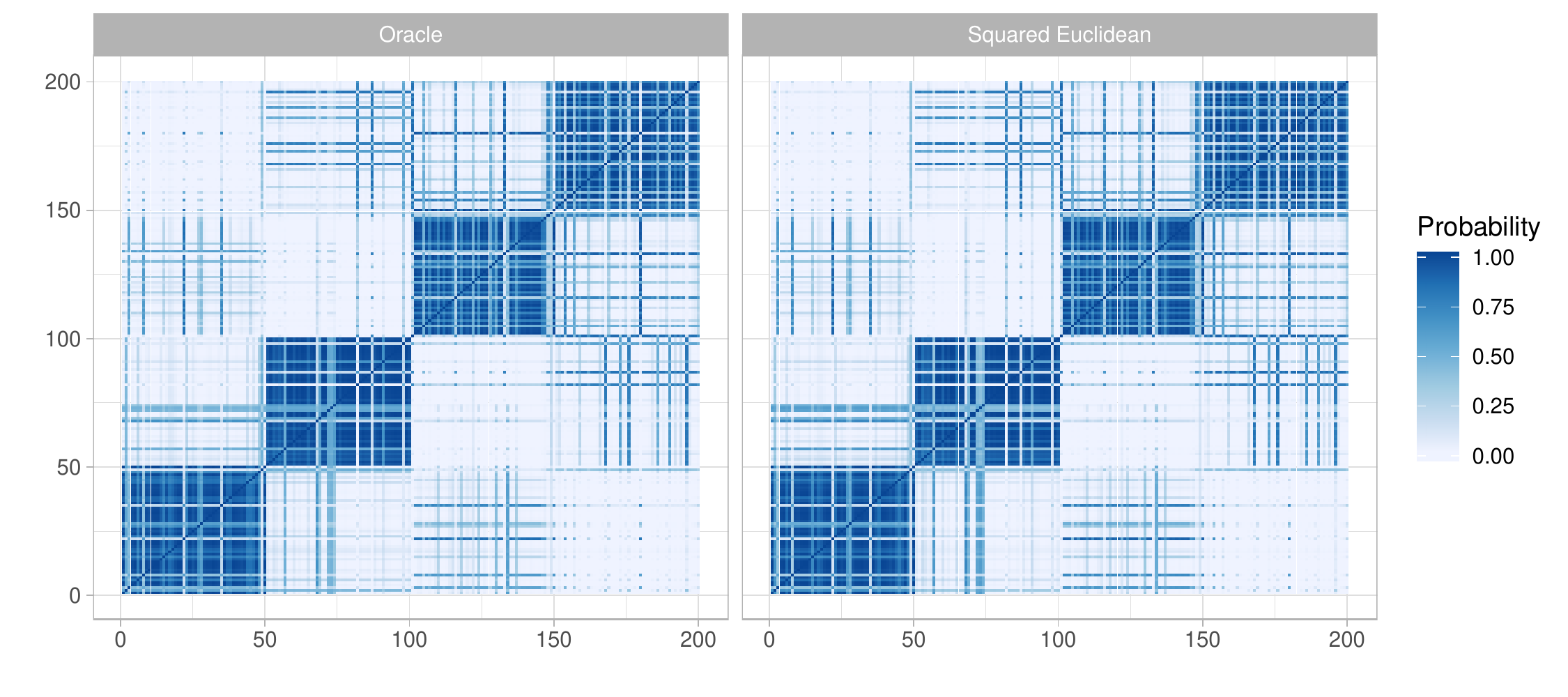}
\caption{Co-clustering probabilities $\bm{S}_\textsc{oracle}$ and $\bm{S}$ for the oracle model and the squared Euclidean \textsc{gb-ppm}. Data points are  generated from equation~\eqref{sim1}, with $\sigma^2 = 1.5$ and ordered according to the true partition $\bm{c}_0$.\label{fig:sim1_co}}
\end{figure}

\begin{figure}[p]
\includegraphics[width=\textwidth]{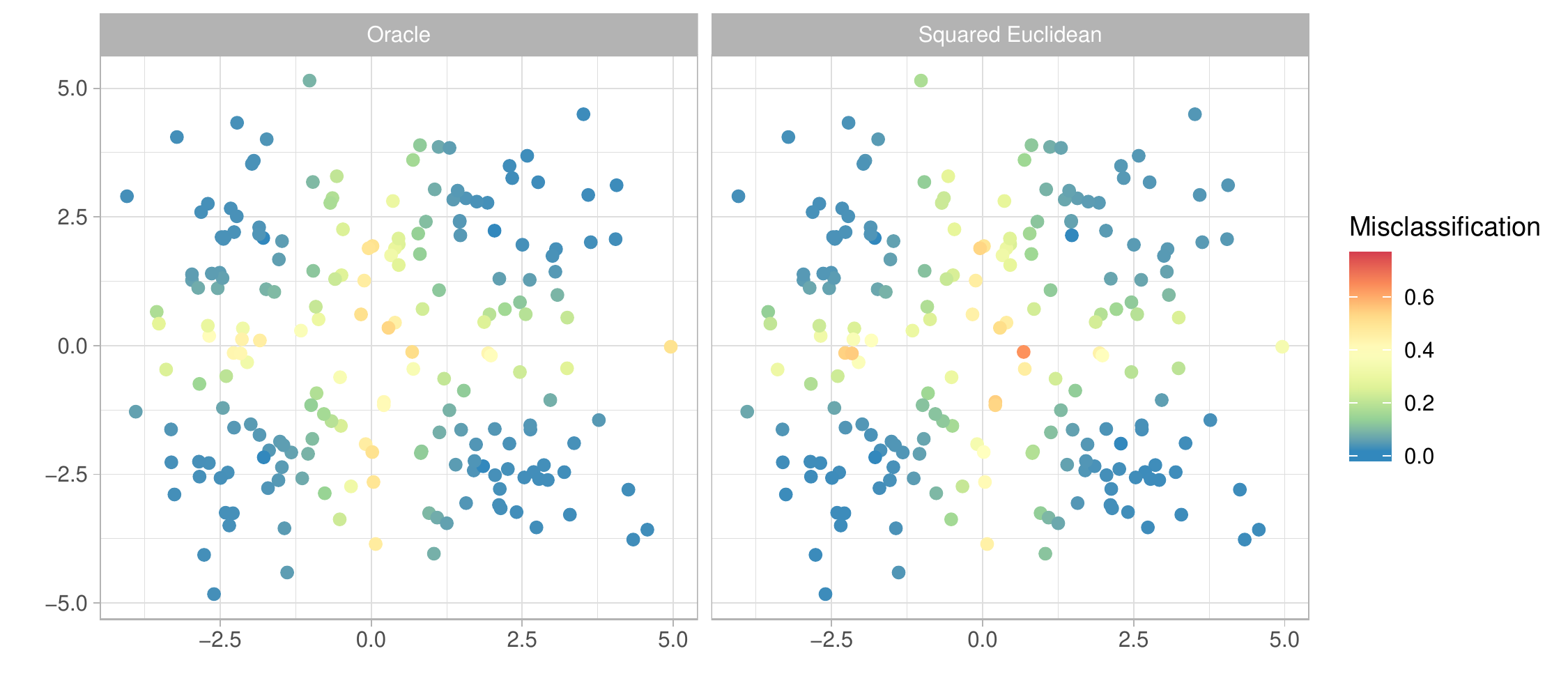}
\caption{Synthetic data points generated from equation~\eqref{sim1} with $\sigma^2 = 1.5$. Colors represent the misclassification probabilities obtained from the oracle co-clustering matrix $\bm{S}_\textsc{oracle}$ and squared Euclidean co-clustering matrix $\bm{S}$, respectively. \label{fig:sim1_miscl}}
\end{figure}

From equation~\eqref{miscl_probabilities} one can obtain \emph{misclassification probabilities} $1 - s_{i \hat{i}^*}$ of a point estimate $\hat{\bm{c}}$, where $\hat{i}^* = i^*_{\hat{c}_i}$ denotes the medoid of the cluster to which the $i$th unit has been allocated. The quantity $1 - s_{i \hat{i}^*}$ approximates the probability that the $i$th unit is allocated to a cluster different from $\hat{c}_i$. To assess the appropriateness of these indicators, we compare them with a gold standard, which we call oracle clustering. Specifically, the oracle distribution associated to the generative mechanism~\eqref{sim1} is
\begin{equation*}
\pi_\textsc{oracle}(\bm{c} \mid \bm{\mu}_1,\dots,\bm{\mu}_K, \sigma^2, \bm{X}) \propto \prod_{i=1}^n \prod_{k=1}^K \mbox{N}(\bm{x}_i \mid \bm{\mu}_k, \sigma^2I_2)^{\mathds{1}(c_i = k)},
\end{equation*}
where $\mathds{1}(\cdot)$ denotes the indicator function, $\bm{\mu}_k$ and $\sigma^2$ are defined as in~\eqref{sim1}, and $\mbox{N}(\bm{x} \mid \bm{\mu}, \bm{\Sigma})$ is the multivariate Gaussian density function. The oracle clustering distribution represents the posterior distribution if the data generating process were known. From $\pi_\textsc{oracle}(\bm{c} \mid \bm{X})$ we derive the corresponding point estimate $\hat{\bm{c}}_{\textsc{oracle}} = \arg \max_{\bm{c}} \pi_\textsc{oracle}(\bm{c} \mid \bm{X})$, the  co-clustering matrix $\bm{S}_\textsc{oracle}$ having entries $s_{ii',\textsc{oracle}}$ for $i,i'=1,\dots,n$, and the associated misclassification probabilities.

\begin{table}
\centering
\begin{tabular}{rrrr}
	\toprule
 & $\sigma^2$ & $\{n(n-1)/2\}^{-1}\sum_{i < i'}|s_{ii'} - s_{ii',\textsc{oracle}}|$ & $d_\textsc{vi}(\hat{\bm{c}}_\textsc{map},\hat{\bm{c}}_{1-\alpha,\textsc{map}})$  \\ 
 	\midrule
  & 0.75 & 0.0092 & 0.5251  \\ 
  & 1.5 & 0.0180 & 1.4215  \\ 
  & 3 & 0.0414 & 2.2243 \\ 
	\bottomrule
\end{tabular}
\caption{Summary of results for the first simulation study. Data points are generated from~\eqref{sim1}, for various values of $\sigma^2$. \label{tab1}}
\end{table}

In Figure~\ref{fig:sim1_co} we graphically compare the co-clustering matrices $\bm{S}$ and $\bm{S}_\textsc{oracle}$, which are almost indistinguishable. A more precise quantification of the discrepancy between these two quantities is given in Table~\ref{tab1}, where we report the average absolute deviations between the entries of $\bm{S}$ and $\bm{S}_\textsc{oracle}$, for a few values of $\sigma^2$. Although the differences increase slightly when $\sigma^2$ is large, this simulation suggests that Gibbs posteriors may be very close to the oracle distribution. In Figure~\ref{fig:sim1_miscl} we show the misclassification probabilities  for the oracle and for our \textsc{gb-ppm} model which are, unsurprisingly, almost identical.

\subsection{Robust clustering}\label{sec:sim2}

In the second simulation study we also consider $n = 200$ observations evenly divided in $K = 4$ clusters, each having $n_1 = \cdots = n_4 = 50$ data points. Within each partition we assume that
\begin{equation}\label{sim2}
(\bm{x}_i \mid \bm{\mu}_k, \sigma^2, c_i = k) \overset{\textup{iid}}{\sim} t_2\left(\bm{\mu}_k, \sigma^2 I_2\right), \qquad i \in C_k, \quad k=1,\dots,K,
\end{equation}
where $t_2(\bm{\mu}, \bm{\Sigma})$ is a multivariate Student's $t$-distribution with location $\bm{\mu}$, scale $\bm{\Sigma}$, and  $2$ degrees of freedom. We let $\bm{\mu}_1 = (-2,-2)$, $\bm{\mu}_2 = (-2,2)$, $\bm{\mu}_3 = (2,-2)$, $\bm{\mu}_4 = (2,2)$ and $\sigma^2 = 1$. The random vectors $\bm{x}_i$ in equation~\eqref{sim2} have finite expectation but infinite variance, meaning that ``outliers'' are expected. We consider two different \textsc{gb-ppm}s, namely the k-means model of Section~\ref{sec:squared_euclidean} and the \textsc{gb-ppm} with Manhattan dissimilarities, which arises when $p=1$ in the class described in Section~\ref{sec:Minkowski}. For uncertainty quantification we estimate $\lambda$ using the default strategies discussed in the aforementioned sections.  Neither model is compatible with the generative process in equation~\eqref{sim2}, but this is not a concern within our generalized Bayes framework.

\begin{figure}[tp]
\includegraphics[width=\textwidth]{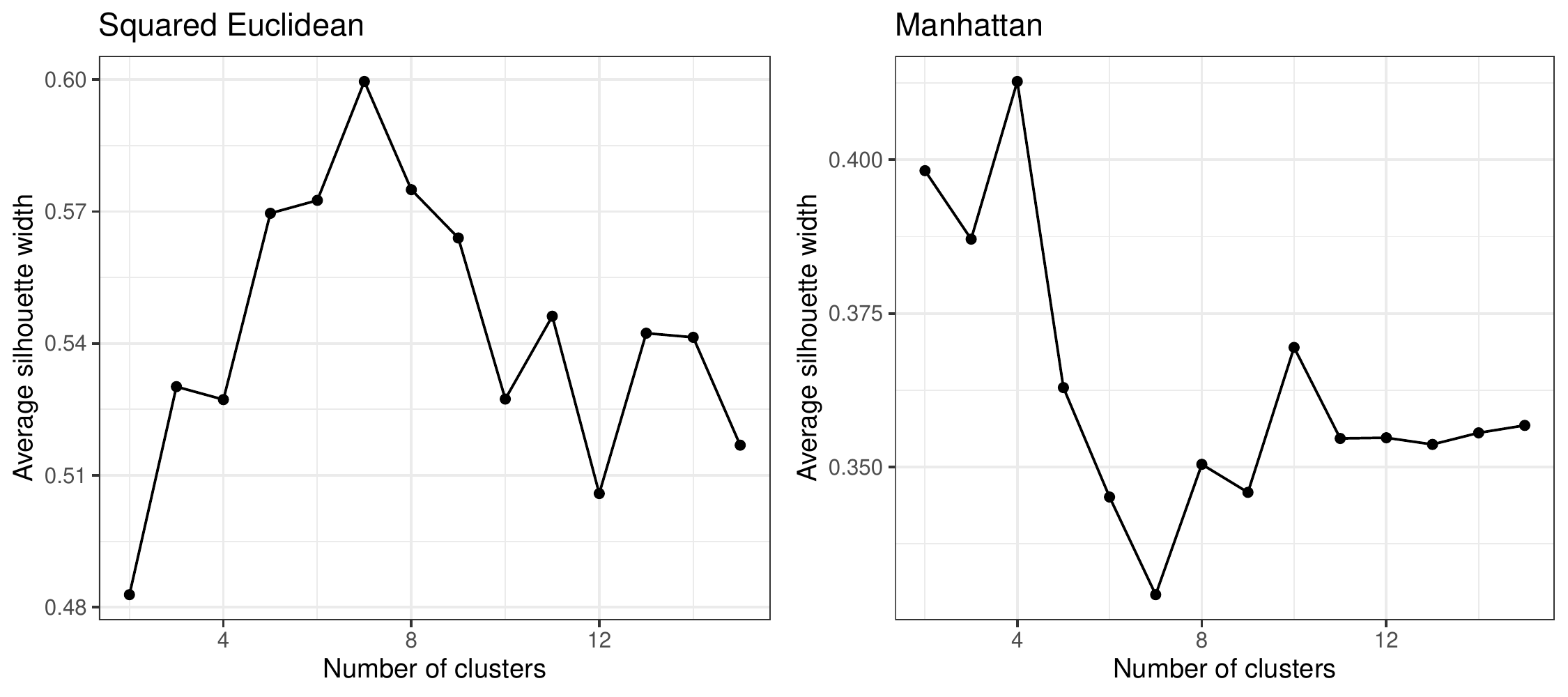}
\caption{Average silhouette width associated to $\hat{\bm{c}}_\textsc{map}$ for different values of $K$, using the k-means algorithm (left plot) and the k-dissimilarities algorithm with Manhattan distances (right plot). Data points are generated according to equation~\eqref{sim2}. \label{fig:sim2_select}}
\end{figure}

Point estimates $\hat{\bm{c}}_\textsc{map}$ are obtained in both cases through the k-dissimilarity algorithm. Consistent with the discussion of Section~\ref{sec:bissiri}, we select the number of clusters $K$ in an exploratory phase before uncertainty quantification. In practice, one can rely on well-known heuristic methods; eg refer to ~\citet{Kaufman1990,Hastie2008}.  We use the average silhouette statistic \citep{Kaufman1990}, which is a widely used goodness-of-fit index for dissimilarity matrices. The results are reported in Figure~\ref{fig:sim2_select}, which shows that the model with Manhattan dissimilarities favors $K = 4$ clusters, whereas this criteria in the squared Euclidean case leads to $K = 7$ clusters. In practice, one may prefer the Manhattan model because it leads to a simpler clustering solution. This marked difference in the silhouette statistics is likely due to the presence of outliers, which will be clear from our quantification of the uncertainty.

\begin{figure}[tp]
\includegraphics[width=\textwidth]{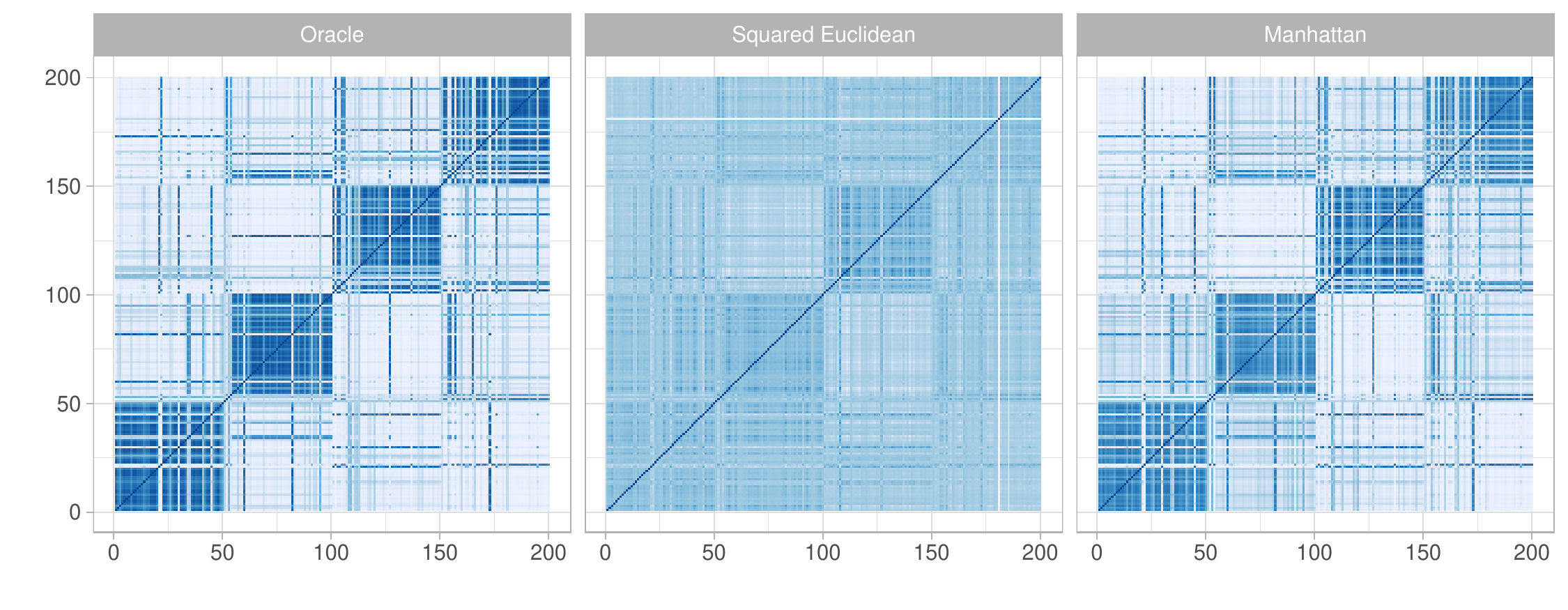}
\caption{Co-clustering probabilities for the oracle model, the squared Euclidean \textsc{gb-ppm}, and the \textsc{gb-ppm} with Manhattan dissimilarities. Colors correspond to probabilities, from white (low probability) to dark blue (high probability). Data points are  generated from equation~\eqref{sim2} and ordered according to the true partition $\bm{c}_0$. \label{fig:sim2_co}}
\end{figure}

To make the results comparable, we fix $K = 4$ in both \textsc{gb-ppm}s and run Gibbs samplers for $5,000$ iterations, having discarded $1,000$ samples as burn-in. The traceplots of the losses show good mixing and no evidence against convergence. We compute the associated co-clustering matrices and the oracle matrix $\bm{S}_\textsc{oracle}$, which is defined as in Section~\ref{sec:sim1}, having replaced the multivariate Gaussian with the multivariate Student's t-distribution. These matrices are displayed in Figure~\ref{fig:sim2_co}.  This graph strongly suggests that the \textsc{gb-ppm} based on Manhattan pairwise dissimilarities outperforms the one relying on the squared Euclidean loss. Indeed, the co-clustering matrix of the former is highly similar to $\bm{S}_\textsc{oracle}$, which is the gold standard. In the k-means case, the presence of outliers leads to highly unreliable uncertainty quantification.  These results are not surprising as absolute deviations have long been used in place of squared losses to robustify clustering; this motivated the  \textsc{pam} algorithm \citep{Kaufman1990}, which is closely related to our \textsc{gb-ppm} with pairwise dissimilarities.

\subsection{The carcinoma dataset}\label{sec:realdata}

In the carcinoma dataset~\citep[][Chap. 13]{Agresti2002}, seven pathologists separately classified $n=118$ slides regarding the presence or absence of carcinoma of the uterine cervix. We are interested in finding groups among these ratings, in order to identify interpretable patterns. In addition, we aim at providing a probabilistic assessment of the cluster allocation if a new slide were examined.  

\begin{figure}[tp]
\centering
\includegraphics[width=0.8\textwidth]{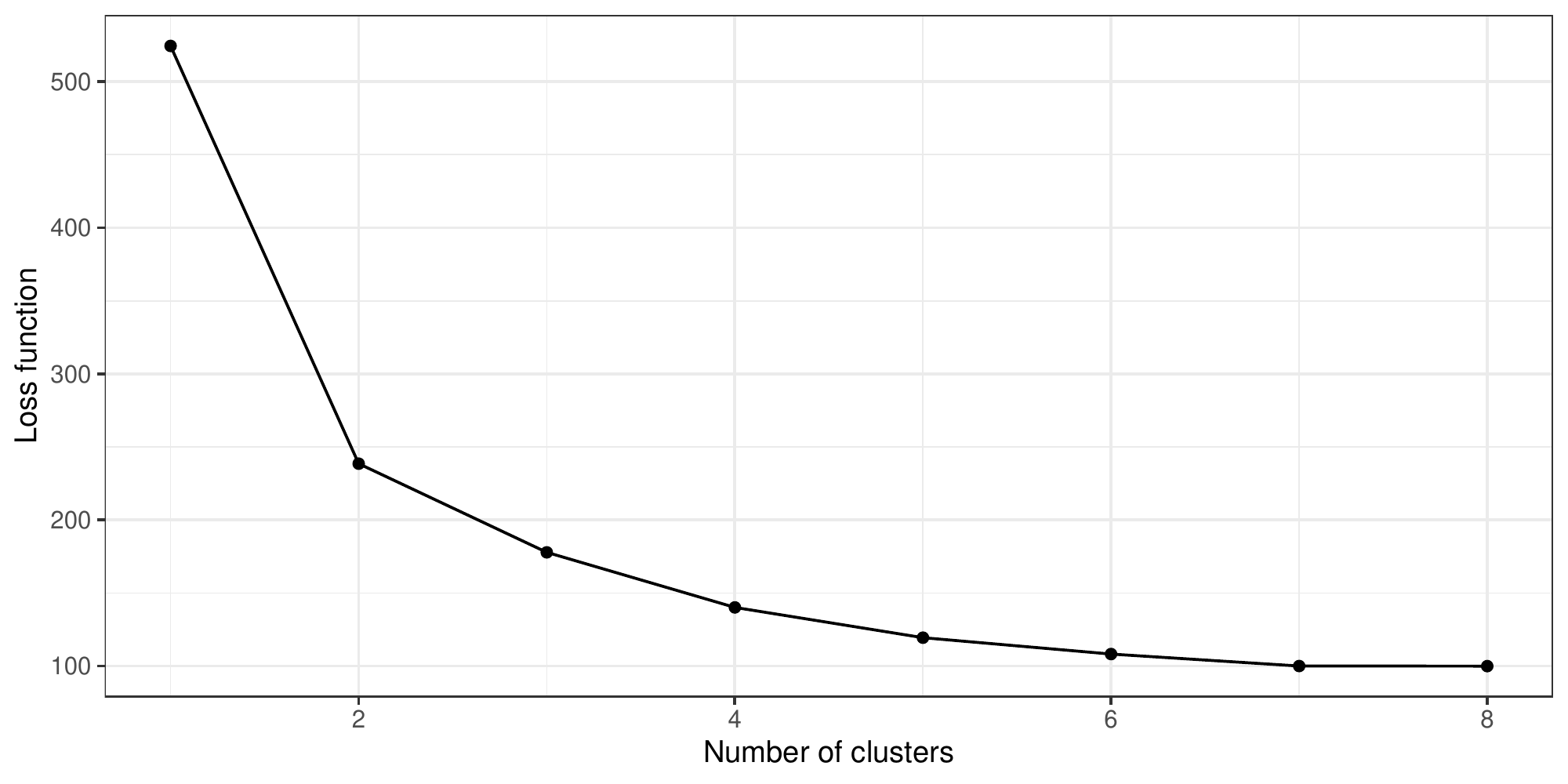}
\caption{Values of the binary loss $\ell(\hat{\bm{c}}_\textsc{map}; \bm{X})$ of equation~\eqref{kl_gbppm} evaluated at the \textsc{map}, for the carcinoma dataset and for different values of $K$. \label{fig:app_select}}
\end{figure}

The dataset $\bm{X}$  comprises $118 \times 7$ binary indicators, implying that the \textsc{gb-ppm} of Section~\ref{sec:kl_model} is a reasonable model choice. We first obtained the $\hat{\bm{c}}_\textsc{map}$ point estimate for different values of $K$, using Algorithm~\ref{Algorithm3}. We report in Figure~\ref{fig:app_select} the sequence of minimized losses, which decreases as a function of $K$. Application of the elbow rule suggests that between $2$ and $4$ clusters are needed to properly summarize the data. We choose $K = 3$, also on the basis of the following qualitative judgement. Specifically, we expect the evaluations of the pathologist to agree in most cases, meaning that each subject should get either mostly negative or mostly positive diagnoses. These cases represent the two main clusters. However, there might be a third group of subjects whose diagnosis is unclear, which indeed represents a refinement with respect to the initial dichotomic scenario. This description will be empirically confirmed in the data, supporting the choice $K=3$.

\begin{figure}[tp]
\includegraphics[width=\textwidth]{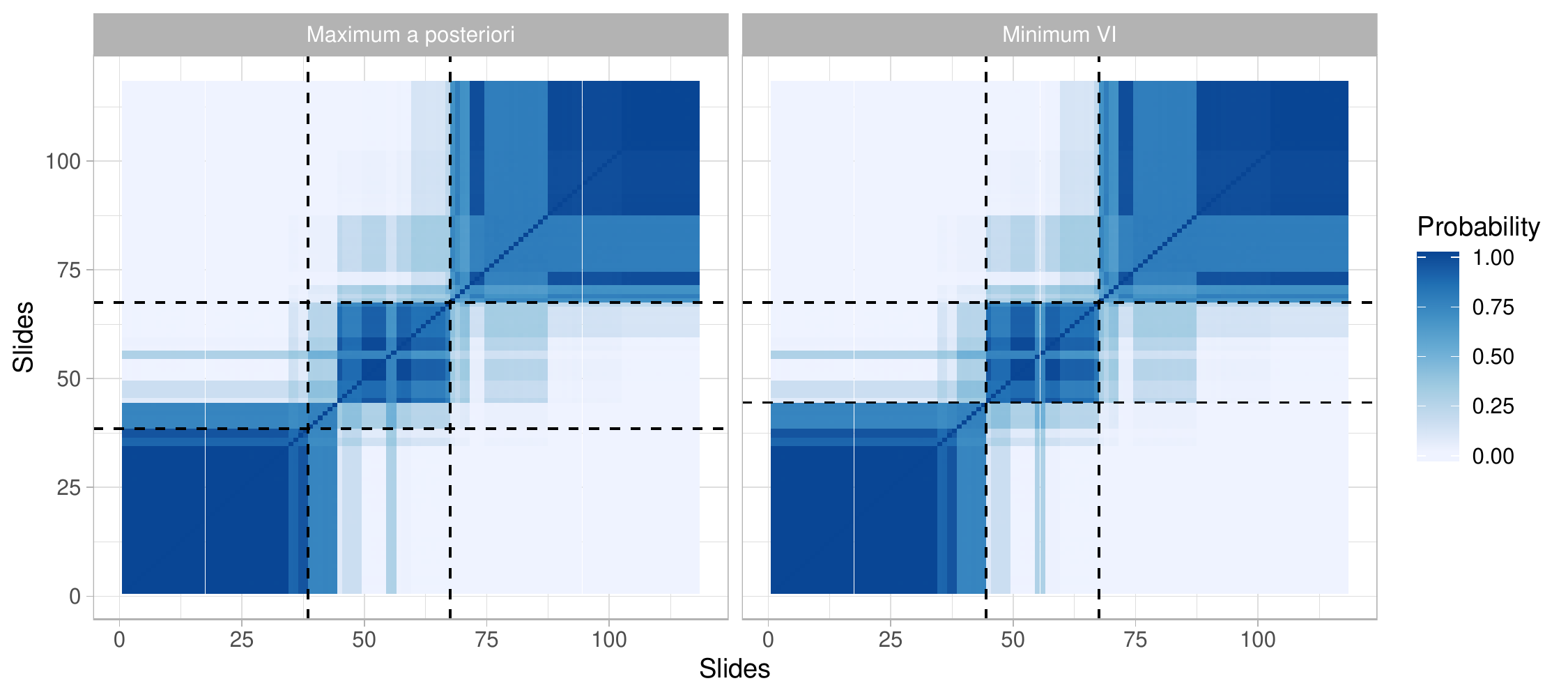}
\caption{Co-clustering probabilities for the Kullback-Leibler \textsc{gb-ppm} applied to the carcinoma dataset. Dashed lines partition the data according to the point estimates $\hat{\bm{c}}_\textsc{map}$ (left plot) and $\hat{\bm{c}}_\textsc{vi}$ (right plot). Observations are ordered in both plots according to the $\hat{\bm{c}}_\textsc{map}$ estimate.  \label{fig:app_co}}
\end{figure}

We run the Gibbs sampler for $16,000$ iterations and we discard the first $1,000$ samples as a burn-in. In the left plot of Figure~\ref{fig:app_co} we depict the co-clustering matrix $\bm{S}$, together with the estimated partition $\hat{\bm{c}}_\textsc{map}$. Although the \textsc{map} is mostly consistent with the pairwise probabilities in~$\bm{S}$, there are six subjects in the second cluster that one would expect to be allocated in the first group according to $\bm{S}$. This is due to the fact that the \textsc{map} is not necessarily a good point estimate if one is interested in ``average'' behaviors, such as those of the co-clustering matrix $\bm{S}$. To overcome this difficulty, we rely on the alternative point estimate $\hat{\bm{c}}_\textsc{vi}$, which is the value minimizing the posterior variation of information distance. As described in \citet{Wade2018}, the estimate $\hat{\bm{c}}_\textsc{vi}$ can be approximated by considering a lower-bound of the posterior expectation of $d_\textsc{vi}(\bm{c},\bm{c}')$. Albeit $\hat{\bm{c}}_\textsc{vi}$ is computationally less convenient than $\hat{\bm{c}}_\textsc{map}$, it indeed provides a better posterior summary, as illustrated in Figure~\ref{fig:app_co}. 

\begin{table}
\centering
\begin{tabular}{rrrrrrrr}
\toprule
Pathologist & \textsc{a} & \textsc{b} & \textsc{c} & \textsc{d} & \textsc{e} & \textsc{f} & \textsc{g} \\ 
\midrule
$\tilde{\bm{x}}_1 $ & 0.06 & 0.14 & 0.01 & 0.01 & 0.06 & 0.01 & 0.01 \\ 
$\tilde{\bm{x}}_2 $ & 0.56 & 0.98 & 0.02 & 0.06 & 0.77 & 0.02 & 0.65 \\ 
$\tilde{\bm{x}}_3 $ & 0.99 & 0.97 & 0.87 & 0.61 & 0.99 & 0.49 & 0.99 \\ 
\bottomrule
\end{tabular}
\caption{Adjusted centroids $\tilde{\bm{x}}_j = (\tilde{x}_{j1},\dots,\tilde{x}_{j7})^\intercal$, $j=1,\dots,3$, of the \textsc{gb-ppm} associated to the variation of information estimate $\hat{\bm{c}}_\textsc{vi}$. \label{tab2}}
\end{table}

\begin{table}
\centering
\begin{tabular}{lrrr}
\toprule
Data points & Cluster 1 & Cluster 2 & Cluster 3 \\ 
\midrule
$\bm{x}_{n+1}=(0,1,0,0,0,0,0)$ & 0.79 & 0.21 & 0.00 \\ 
$\bm{x}_{n+2}=(0,1,0,0,1,0,0)$ & 0.06 & 0.94 & 0.00 \\ 
$\bm{x}_{n+3}=(1,1,1,0,1,0,1)$& 0.00 & 0.04 & 0.96 \\ 
\bottomrule
\end{tabular}
\caption{Predictive probabilities for the data points $\bm{x}_{n+1}$, $\bm{x}_{n+2}$, and $\bm{x}_{n+3}$ calculated as in~\eqref{cluster_predictive}, using the $\hat{\bm{c}}_\textsc{vi}$ point estimate as reference. \label{tab3}}
\end{table}

In Table~\ref{tab2} we report the adjusted centroids associated to the point estimate $\hat{\bm{c}}_\textsc{vi}$. Interestingly, they have a clear and simple interpretation. The first cluster refers to cases where all pathologists agree there is no carcinoma, with the occasional exception of $\textsc{b}$. Conversely, the third cluster refers to cases where most pathologists (\textsc{a}, \textsc{b}, \textsc{c}, \textsc{e}, and \textsc{g}) agree with its presence. Finally, the second cluster refer to cases where there is disagreement among the doctors. This analysis is consistent with the finding of \citet{Agresti2002} and supports the choice $K = 3$. For illustrative purposes, we also report in Table~\ref{tab3} the predictive allocation probabilities associated to $3$ new potential data points. For example, consistently with the above discussion, when the pathologist \textsc{a}, \textsc{b}, \textsc{c}, \textsc{e}, and \textsc{g} agree about the presence of the carciroma, the probability of this new data point being allocated to the third cluster is about $0.96$.

\section{Discussion}\label{sec:discussion}

In this paper we presented a wide class of generalized Bayes models based on Gibbs posteriors termed \textsc{gb-ppm}. We studied its properties, proposed novel inferential routines and suggested practical usages. Our modeling overcome many limitations of standard Bayesian mixtures, since it leads to computationally efficient algorithms and robust specifications while allowing uncertainty quantification. In addition, our approach merges loss- and model-based approaches into a unified inferential framework. However, the aforementioned advantages are not evenly shared by all \textsc{gb-ppm}s. Indeed, each loss has its own peculiarities, and it favors the formation of specific clusters' shapes. Thus, the loss function should be carefully selected on the basis of the applications' aims. Note that our probabilistic interpretation of the Bregman and pairwise dissimilarity cohesions could be very helpful in this sense. This reasoning is consistent with the notion of ``optimal'' partition $\bm{c}_\textsc{opt}$ described in Section~\ref{sec:bissiri}, which therefore should not be interpreted as optimal in absolute terms but only in reference to the specific application.  Although the sensitivity to the loss' choice parallels the difficulty in the kernel's choice in standard mixture models, note that i) generalized Bayes models might have crucial computational advantages, ii) the choice of the loss is often easier and conceptually simpler than the elicitation of a probabilistic model, and iii) losses can be regarded as a generalization of likelihoods, therefore enlarging the modeling possibilities rather than representing an alternative framework. Finally, we note that the choice of the $\lambda$ parameter remains partially unaddressed beyond the two broad classes of losses described in this paper. Although few general strategies are discussed e.g. in \citet{Bissiri2016, Holmes2017}, their concrete application to clustering problems is unclear. We believe this issue is an interesting research direction for future works.

\section*{Acknowledgements} This work was partially supported by grants R01ES027498 and R01ES028804 of the National Institute of Environmental Health Sciences of the United States National Institutes of Health.

\appendix

\section{Computational details for a \textsc{gb-ppm} with average dissimilarities}\label{app1}

In a \textsc{gb-ppm} with average dissimilarities, a recursive formula for the the differences between the discrepancies is available. The latter appears in Algorithm~\ref{Algorithm4} and in Gibbs sampling via Theorem~\ref{GibbsSampling}. This leads to considerable computational improvements, because one does not need to re-compute the discrepancies $\mathcal{D}_\gamma(\bm{x}_{i'}; \bm{X}_k)$ at every step of the algorithms.  Specifically, let $\mathcal{D}_\gamma(\bm{x}_i; \bm{X}_k) = 1/n_k \sum_{i' \in C_k} \gamma(||\bm{x}_i - \bm{x}_{i'}||_p^p)$ be the average dissimilarity as in equation~\eqref{average}. Then,
\begin{equation*}
\sum_{i' \in C_k} \mathcal{D}_\gamma(\bm{x}_{i'}; \bm{X}_k) - \sum_{i' \in C_{k,-i}} \mathcal{D}_\gamma(\bm{x}_{i'}; \bm{X}_{k,-i}) = \frac{2}{n_k}\sum_{i' \in C_k} \gamma(||\bm{x}_i - \bm{x}_{i'}||_p^p) - \frac{1}{n_k}\sum_{i' \in C_{k,-i}}\mathcal{D}_\gamma(\bm{x}_{i'}; \bm{X}_{k,-i}).
\end{equation*}
Hence, at each step one can update the real numbers $\sum_{i' \in C_k} \mathcal{D}_\gamma(\bm{x}_{i'}; \bm{X}_k)$, $\sum_{i' \in C_{k,-i}} \mathcal{D}_\gamma(\bm{x}_{i'}; \bm{X}_{k,-i})$ by simply adjusting for the factor $2/n_k\sum_{i' \in C_k} \gamma(||\bm{x}_i - \bm{x}_{i'}||_p^p)$, which is easy to compute. 

\section{Proofs}

\subsection*{Proof of Theorem~\ref{GibbsSampling}}

By definition, the desired full conditional distribution is equal to $\pi(c_i \mid \bm{c}_{-i},\bm{X}) \propto \pi(\bm{c} \mid \lambda, \bm{X})$. Therefore for any $\bm{c} : |\bm{C}| = K$, the probability of $c_i$ being allocated to one of the previous values is
\begin{equation*}
\begin{aligned}
\mathds{P}(c_i = k \mid \bm{c}_{-i}, \lambda, \bm{X}) &\propto \prod_{j\neq k} \rho(C_{j,-i};\lambda, \bm{X}_{j,-i}) \rho(C_k,\lambda, \bm{X}_k), \qquad k=1, \dots, K. \\
\end{aligned}
\end{equation*}
Then, multiply the above term by  $1/\prod_{k=1}^{K} \rho(C_{k,-i};\lambda, \bm{X}_{k,-i})$, so that
\begin{equation*}
\begin{aligned}
\mathds{P}(c_i = k \mid \bm{c}_{-i}, \lambda, \bm{X}) & \propto \frac{\rho(C_k,\lambda, \bm{X}_k)}{\rho(C_{k,-i},\lambda, \bm{X}_{k,-i})}, \qquad k=1, \dots,K,
\end{aligned}
\end{equation*}
and the result follows. 

\subsection*{Proof of Theorem~\ref{profile_lik}}

We begin by stating the following Lemma, whose proof has been omitted since it can be found in \citet{Banerjee2005} in the $\lambda = 1$ case. The extension to a general $\lambda$ is trivial. 

\begin{lemma} Let $\pi_{\textsc{ed}}(\bm{x} \mid \bm{\theta},\lambda) = \pi(\bm{x} \mid \lambda) \exp\{\lambda [\bm{\theta}^\intercal\bm{x} - \kappa(\bm{\theta})]\}$ belong to an exponential dispersion family. Then there exists a function $\varphi(\bm{x})$ such that
\begin{equation*}
\lambda [\bm{\theta}^\intercal\bm{x} - \kappa(\bm{\theta})] = - \lambda \mathcal{D}_\varphi(\bm{x}; \mu(\bm{\theta})) + \lambda \varphi(\bm{x}),
\end{equation*}
where $\mu(\bm{\theta}) =  \int_\mathds{X} \bm{x} e^{\lambda [\bm{\theta}^\intercal\bm{x} - \kappa(\bm{\theta})]}\Pi_\lambda(\dd \bm{x})$. 
\end{lemma}

Therefore, the posterior distribution in a mixture model with exponential dispersion components can be written as
\begin{equation*}
\begin{aligned}
\pi_\textsc{ed}(\bm{c} \mid \lambda, \bm{\theta}_1,\dots,\bm{\theta}_K, \bm{X}) &\propto \prod_{k=1}^K  \prod_{i \in C_k} \pi_{\textsc{ed}}(\bm{x}_i \mid \bm{\theta}_k, \lambda) \\
&\propto \prod_{k=1}^K  \prod_{i \in C_k} \pi(\bm{x}_i \mid \lambda) \exp\left\{\lambda [\bm{\theta}_k^\intercal\bm{x}_i - \kappa(\bm{\theta}_k)]\right\} \\
&\propto \prod_{k=1}^K  \left[\prod_{i \in C_k} \pi(\bm{x}_i \mid \lambda) \exp\{\lambda \varphi(\bm{x}_i)\}\right] \exp\left\{- \lambda\sum_{i \in C_k}  \mathcal{D}_\varphi[\bm{x}_i; \mu(\bm{\theta}_k)]\right\} \\
&\propto \prod_{k=1}^K  \exp\left\{- \lambda\sum_{i \in C_k}  \mathcal{D}_\varphi[\bm{x}_i; \mu(\bm{\theta}_k)]\right\},
\end{aligned}
\end{equation*}
which holds for any value of $\bm{\theta}_1,\dots,\bm{\theta}_K$. Then, the maximum likelihood estimate for each $\bm{\theta}_k$ can be computed separately, since the likelihood factorizes. Thus, we seek
\begin{equation*}
\hat{\bm{\theta}}_k = \arg \max_{\bm{\theta}}  \prod_{i \in C_k} \pi_{\textsc{ed}}(\bm{x}_i \mid \bm{\theta}, \lambda) = \arg \max_{\bm{\theta}} \sum_{i \in C_k} \left[ \bm{\theta}^\intercal\bm{x}_i - \kappa(\bm{\theta})\right], \qquad k=1,\dots,K,
\end{equation*}
if they exist. By differentiating the right-hand side of the above equation and equating to zero, we obtain that $\hat{\bm{\theta}}_k$ is the solution of
\begin{equation*}
\bar{\bm{x}}_k = \nabla \kappa(\hat{\bm{\theta}}_k) = \mu(\hat{\bm{\theta}}_k),
\end{equation*}
where the last equality follows from the property of exponential dispersion families \citep{Jorgensen1987}. Therefore, we have $\hat{\bm{\theta}}_k = \mu^{-1}(\bar{\bm{x}}_k) = \theta(\bar{\bm{x}}_k)$, which concludes the proof.

\subsection*{Proof of Proposition~\ref{monotone}}
Broadly speaking, the proof follows because we alternate between two minimization procedures. Let us define
\begin{equation*}
\ell(\bm{c}, \bm{\theta}_1,\dots,\bm{\theta}_K; \bm{X}) = \sum_{k=1}^K\left[ -\log \pi(\bm{\theta}_k \mid \lambda) + \lambda \sum_{i \in C_k} \mathcal{D}_\varphi(\bm{x}_i; \mu(\bm{\theta}_k))\right],
\end{equation*}
and recall that
\begin{equation*}
\ell(\bm{c}; \bm{X}) = \min_{\bm{\theta}_1,\dots,\bm{\theta}_K} \ell(\bm{c}, \bm{\theta}_1,\dots,\bm{\theta}_K; \bm{X}).
\end{equation*}
Let $\bm{c}^{(t)}$ be the partition at the $t$th iteration and let $\bm{m}_1^{(t)},\dots,\bm{m}_K^{(t)}$ be the corresponding centroids. Then,
\begin{equation*}
\begin{aligned}
 \ell(\bm{c}^{(t)}; \bm{X}) = \ell(\bm{c}^{(t)}, \bm{m}_1^{(t)},\dots,\bm{m}_K^{(t)}; \bm{X}) & \ge \ell(\bm{c}^{(t+1)}, \bm{m}_1^{(t)},\dots,\bm{m}_K^{(t)}; \bm{X}) \\
 & \ge \ell(\bm{c}^{(t+1)}, \bm{m}_1^{(t+1)},\dots,\bm{m}_K^{(t+1)}; \bm{X}) =  \ell(\bm{c}^{(t+1)}; \bm{X}).
\end{aligned} 
\end{equation*}
The first inequality follows because the re-allocation step minimizes the associated Bregman divergences, whereas the second follows because the adjusted centroids are obtained as the solution of a minimization problem. Moreover, the partition space is finite and this, together with monotonicity, implies that the number of steps will be finite. 

\subsection*{Proof of Proposition~\ref{monotone2}}
Broadly speaking, the proof follows because Algorithm~\ref{Algorithm4} iterates over $n$ minimization procedures. More precisely, let $\bm{c}^{(t)} = (c_1^{(t)},\dots,c_n^{(t)})$ be the partition at the $t$th iteration. Then,
\begin{equation*}
\begin{aligned}
 \ell(\bm{c}^{(t)}; \bm{X}) = \ell(c_1^{(t)},\dots,c_n^{(t)}; \bm{X}) & \ge \ell(c_1^{(t+1)},\dots,c_n^{(t)}; \bm{X}) \ge \cdots \ge  \ell(c_1^{(t+1)},\dots,c_n^{(t+1)}; \bm{X}) =  \ell(\bm{c}^{(t+1)}; \bm{X}).
\end{aligned} 
\end{equation*}
Moreover, the partition space is finite and this, together with monotonicity, implies that the number of steps will be finite. 

\subsection*{Proof of Theorem~\ref{pairwise_lik}}

The proof of Theorem~\ref{pairwise_lik} follows from the properties of $L^p$ spherical distributions. Indeed, any $L^p$ spherical distribution  admits a useful stochastic representation. More precisely, a random vector $\bm{x} \sim \pi_\textsc{sp}(\bm{x})$ is $L^p$ spherical if and only if it can be written as $\bm{x} \overset{\text{d}}{=} r \bm{u}$, where $r > 0$ is a positive random variable which is independent of the random vector $\bm{u} \in \mathds{R}^d$, which follows the $L^p$ uniform distribution \citep{Gupta1997}. The random variable $r \sim \pi_\textsc{r}(r)$ is the \emph{radius}, because if $\bm{x} \sim \pi_\textsc{sp}(\bm{x})$ and $\bm{x} \overset{\text{d}}{=} r \bm{u}$, then $r \overset{\text{d}}{=} ||\bm{x}||_p$. In addition, the density of the radius $\pi_\textsc{r}(r)$ characterizes the density  of the whole vector $\pi_\textsc{sp}(\bm{x})$ and vice versa, namely
\begin{equation*}\label{Lp_spherical}
\pi_\textsc{sp}(\bm{x}) = g(||\bm{x}||_p^p)=  b_{d,p} \: \pi_\textsc{r}(||\bm{x}||_p)||\bm{x}||_p^{1-d}, \qquad \pi_\textsc{r}(r) = \frac{1}{b_{d,p}} r^{d-1} g(r^p),
\end{equation*}
for any $r > 0$ and $\bm{x} \in \mathds{R}^d$, where $b_{d,p} = p^{d-1} \Gamma(d/p)/[2 \Gamma(1/p)]^d$ is a normalizing constant with $\Gamma(\cdot)$ the gamma function; see \citet{Gupta1997} for details. Therefore, the term $\exp\left\{ -\lambda/2\gamma(||\bm{x}_i - \bm{x}_{i'}||_p^p)\right\}$, appearing in Definition~\ref{pairwise}, is proportional to a proper density of an $L^p$ spherical distribution in $\mathds{R}^d$ if and only if the law of the radius  $\pi_\textsc{r}$ is well defined. This is equivalent to the required condition $\int_{\mathds{R}_+} r^{d-1} \exp\left\{-\lambda/2 \gamma(r^p)\right\}\dd r < \infty$.

\bibliographystyle{chicago}
\bibliography{biblio}

\end{document}